\documentclass[preprint]{aastex}

\newcommand{\ie}{{\frenchspacing\it i.e.}}
\newcommand{\eg}{{\frenchspacing\it e.g.}}

\newcommand{\etal}{{\frenchspacing\it et al. }}

\newcommand{\lsim}{\hbox{ \rlap{\raise 0.425ex\hbox{$<$}}\lower 0.65ex\hbox{$\sim$} }}
\newcommand{\gsim}{\hbox{ \rlap{\raise 0.425ex\hbox{$>$}}\lower 0.65ex\hbox{$\sim$} }}

\shorttitle{Northern Sky Optical Cluster Survey}
\shortauthors{Gal \etal}

\begin{document}

\title{The Northern Sky Optical Cluster Survey II: \\
 An Objective Cluster Catalog for 5800 Square Degrees}

\author{R.R. Gal\altaffilmark{1}, R.R. de Carvalho\altaffilmark{2,3}, P.A.A. Lopes\altaffilmark{2}, S.G. Djorgovski, R.J. Brunner\altaffilmark{4}, A. Mahabal, S.C. Odewahn\altaffilmark{5}}
\affil{Palomar Observatory, Caltech, MC105-24, Pasadena, CA 91125}
\altaffiltext{1}{Center for Astrophysical Sciences, Johns Hopkins University, Baltimore, MD 21218 \\ 
{\indent Email: rrg@pha.jhu.edu} }
\altaffiltext{2}{Observat\'orio Nacional, Rua Gal.  Jos\'e Cristino, 77 -- 20921-400, Rio de Janeiro, RJ, Brazil}
\altaffiltext{3}{Currently at Osservatorio Astronomico di Brera, Via Brera 28, 20121 - Milano, Italy}
\altaffiltext{4}{University of Illinois, Dept. of Astronomy, 1002 W. Green St., Urbana, IL 61801}
\altaffiltext{5}{Arizona State University, Dept. of Physics \& Astronomy, Tempe, AZ 85287} 

\begin{abstract}
We present a new, objectively defined catalog of candidate galaxy clusters
based on the galaxy catalogs from the Digitized Second Palomar Observatory Sky Survey (DPOSS). This cluster catalog, derived from the best calibrated plates in the high latitude ($|b|>30^{\circ}$) Northern Galactic Cap region, covers 5,800 square degrees, and contains 8,155 candidate clusters. A simple adaptive kernel density mapping technique, combined with the SExtractor object detection algorithm, is used to detect galaxy overdensities, which we identify as clusters. Simulations of the background galaxy distribution and clusters of varying richnesses and redshifts allow us to optimize detection parameters, and measure the completeness and contamination rates for our catalog. Cluster richnesses and photometric redshifts are measured, using integrated colors and magnitudes for each cluster. An extensive spectroscopic survey is used to confirm the photometric results. This catalog, with well-characterized sample properties, provides a sound basis for future studies of cluster physics and large scale structure.
\end{abstract}


\section{Introduction}

Over the past two decades, large numbers of astronomers have expended a great deal of effort on the study of galaxy clusters. Motivations for these works cover a wide range of astrophysical topics, from the study of galaxy evolution in dense environments, through the search for dark matter, the measurement of the cluster mass function and its comparison to theoretical predictions, and on to the characterization of large scale structure in the universe. Despite this labor, understanding the underlying physics has often proved enigmatic, due to the complex physical nature of clusters, which is only compounded by the lack of an objectively selected, well-characterized statistical sample. 

To help remedy this situation, and provide a basis for future studies, we have undertaken the creation of a new cluster catalog which fulfills a number of fundamental criteria:
\begin{enumerate}
\item Cluster detection should be performed by an objective, automated algorithm to minimize human biases and fatigue.
\item The algorithm utilized should impose minimal constraints on the physical properties of the clusters, to avoid selection biases.
\item The sample selection function must be well-understood, in terms of both completeness and contamination, as a function of both redshift and richness. The effects of varying the cluster model on the determination of these functions must also be known.
\item The catalog should provide basic physical properties for all the detected clusters, such that specific subsamples can be selected for future study.
\end{enumerate}

In this paper, we describe how we generate a cluster sample that meets the above four criteria, and provide the final catalog for $5800\Box^{\circ}$. We discuss past cluster surveys and their limitations in \S2. In \S3 we briefly review the DPOSS survey \citep{djo03} which provides the basis for our catalog, while \S4 describes the cluster detection technique (a modified version of that presented in \citet{gal00}, hereafter Paper I). \S5 describes simulations used both to optimize the detection algorithm, and define its statistical properties, including selection functions in richness and redshift for each plate. The results of these simulations are compared to extensive spectroscopic follow-up observations. Our photometric redshift estimator and richness measure are described in \S6. An overview of the final catalog is provided in \S7. We conclude with a brief discussion of future extensions of this survey (to cover the South Galactic Cap area, and the remaining areas of the NGP), and discuss some projects we have undertaken utilizing this sample. We use a cosmology with $H_0=67$ km s$^{-1}$ Mpc$^{-1}$ and $q_0=0.5$ throughout; other choices for these parameters have no significant effects on our results except for the poorest and most distant clusters.

\section{A Brief History of Cluster Surveys}

Surveys for galaxy clusters have only recently benefited from the automation afforded by computers. The earliest surveys relied on visual inspection of vast numbers of photographic plates, usually by a single astronomer. As early as 1938, Zwicky discussed such a survey based on plates from the $18''$ Schmidt telescope at Palomar, and results were presented in 1942 by both Zwicky and Katz \& Mulders. Even then, Zwicky, with his typical prescience, noted that elliptical galaxies are much more strongly clustered than late-type galaxies. However, the true pioneering work in this field did not come until 1957, upon the publication of a catalog of galaxy clusters produced by George Abell as his Caltech Ph.D. thesis, which appeared in the literature the following year \citep{abe58}. Zwicky and collaborators followed suit a decade later, with their voluminous Catalogue of Galaxies and of Clusters of Galaxies \citep{zwi68}.

Abell used the red plates of the first National Geographic-Palomar Observatory Sky Survey. In selecting clusters, Abell applied a number of criteria in an attempt to produce a fairly homogeneous catalog. He required a minimum number of galaxies within two magnitudes of the third brightest galaxy in a cluster, a fixed physical area within which galaxies were to be counted, a maximum and minimum distance (redshift) to the clusters, and a minimum galactic latitude to avoid obscuration by interstellar dust. The resulting catalog, consisting of 1,682 clusters in the statistical sample, remained the predominant such resource until 1989. In that year, Abell, Corwin \& Olowin (hereafter ACO) published an improved and expanded catalog, now including the Southern sky. These catalogs have been the foundation for many cosmological studies over the last four decades, despite serious questions about their reliability (which are addressed later in this section). Some other catalogs based on plate material have also been produced, such as \citet{she85}, from the galaxy counts of \citet{sha54}, and a search for more distant clusters carried out on plates from the Palomar $200''$ by \citet{gun86}.

Only in the past ten years has it become possible to utilize the objectivity of computational algorithms in the search for galaxy clusters. These more modern studies required that plates be digitized, so that the data are in machine readable form. Alternatively, the data had to be digital in origin, coming from CCD cameras. Unfortunately, this latter option provided only small area coverage, so the hybrid technology of digitized plate surveys blossomed into a cottage industry, with numerous catalogs being produced in the past decade. All such catalogs relied on two fundamental data sets: the Southern Sky Survey plates, scanned with the Automatic Plate Measuring (APM) machine \citep{mad90} or COSMOS scanner (to produce the Edinburgh/Durham Southern Galaxy Catalog / EDSGC, \citealt{hey89}), and the POSS-I, scanned by the APS group \citep{pen93}.  The first objective catalog produced was the Edinburgh/Durham Cluster Catalog (EDCC, \citealt{lum92}), which covered 0.5 steradians ($\sim 1,600$ square degrees) around the South Galactic Pole (SGP). Later, the APM cluster catalog was created by applying Abell-like criteria to select overdensities from the galaxy catalogs, and is discussed in detail in \citet{dal97}. The work by \citet{ode95}, based on the POSS-I,  provided a Northern sky example of such a catalog, while utilizing additional information (namely galaxy morphology). Some initial work on this problem, using higher quality POSS-II data, was performed by \citet{pic91} in his thesis. 

In addition to these hybrid photo-digital surveys, smaller areas, to much higher redshift, have been covered by numerous deep CCD imaging surveys. Notable examples include the Palomar Distant Cluster Survey (PDCS, \citealt{pos96}), the ESO Imaging Survey (EIS, \citealt{ols99}, \citealt{lob00}), \citet{gon01}, KPNO/Deeprange (\citealt{pos02}), and many others. None of these surveys provide the angular coverage necessary for large-scale structure and cosmology studies, and are typically designed to find rich clusters at high redshift. Only the Sloan Digital Sky Survey (SDSS, \citealt{yor00}) will provide large area, moderately deep CCD coverage. Cluster surveys from the SDSS, including those of \citet{ann99}, \citet{kim02} and \citet{got02}, are only now appearing, and  will cover a much smaller area until the photometric survey is completed.
 
Despite these efforts, one thing is still missing: a catalog of galaxy clusters, produced by objective means, that is at least as deep as the Southern surveys, but which covers the Northern sky. This paper is intended to provide such a catalog.

We note that there have also been many surveys for galaxy clusters at other wavelengths, most notably in the X-ray. All-sky surveys, such as RASS \citep{col00,ebe00,boh00} and EMSS \citep{gio94}, as well as pointed ROSAT observations \citep{sch97,rom00}, have been used to produce cluster catalogs at X-ray wavelengths. Future Sunyaev- Zel'dovich effect (SZ) surveys will also generate extremely important datasets. However, the relationship between optically and X-ray selected clusters is still not fully understood, so an optically selected catalog is essential.

\subsection {Limitations of Existing Catalogs}
 
Most of the optical studies to date have been limited by the statistical 
quality of the available cluster samples. For instance, the Abell catalog
 suffers from a non-objective selection process, poorer plate material, a bias toward centrally-concentrated clusters (especially those with cD galaxies), a relatively low redshift cutoff  ($z \sim0.15$; \citealt{bah83}), and strong plate-to-plate sensitivity variations. Still, many far-reaching  cosmological conclusions have been drawn from it (\ie~ \citealt{bah92}), although later studies have sometimes shown these to be flawed. Photometric errors and other inhomogeneities in the Abell catalog \citep{sut88,efs92}, as well as projection effects \citep{luc83,kat96} are serious and difficult-to-quantify issues. These effects have resulted in discrepant results on the correlation function \citep{bah83,dal97,mil99}, and later attempts to disentangle these issues relied on models to decontaminate the catalog \citep{sut88,oli90}. Also, \citet{sut91} find a 
discrepancy between the angular and spatial correlation function of the Abell
catalog, which is not found in the APM catalog \citep{dal97}. The extent of these effects is also surprisingly unknown; measures of completeness and contamination in the Abell catalog disagree by factors of a few. For instance, \citet{mil99} claim that under- or over-estimation of richness is not a significant problem, whereas \citet{van97} suggest that one-third of Abell clusters have incorrect richnesses, and that one-third of rich ($R\ge1$) clusters are missed.The largest study of $R\ge1$ clusters \citep{mil99} suggests projection effects are not of great concern for the Abell catalog; however, completeness cannot be gauged without deeper samples. Unfortunately, some of these problems will plague any optically selected cluster sample, including our own, but objective selection criteria and a strong statistical understanding of the catalog can mitigate their effects. 

Even some of the other objective catalogs preceding ours have their drawbacks. The APM group, for instance, used digitized $J$ (blue) plates from the Southern Sky Survey; the use of a single blue band provides no color information to distinguish galaxy types, and is a poor choice for cluster detection because clusters are better delineated by redder, early-type galaxies in the redshift range probed ($z<0.3$). Comparable surveys, such as the EDCC, already find a factor of two higher space density of clusters than Abell, and more sensitive CCD surveys find as many as five times more, although these results may be due to strong detection efficiency differences at lower richnesses. Other more recent surveys, such as the EDCCII \citep{bra00} have not yet achieved the area coverage of DPOSS. Additionally, the survey presented here utilizes at least one color (two filters) for photometric redshifts, and a significantly increased amount of CCD photometric calibration data.

\section{DPOSS: A Brief Overview}
The POSS-II \citep{rei91} covers the entire northern sky ($\delta > -3^\circ$) with 897
overlapping fields (each $6.5^\circ$ square, with $5^\circ$ spacings), and,
unlike the old POSS-I, has no gaps in the coverage.  Approximately half
of the survey area is covered at least twice in each band, due to plate
overlaps.  Plates are taken in three bands:
blue-green, IIIa-$J$ + GG395, $\lambda_{\rm eff} \sim480nm$;
red, IIIa-$F$ + RG610, $\lambda_{\rm eff} \sim650nm$; and
very near-IR, IV-$N$ + RG9, $\lambda_{\rm eff} \sim850nm$.
Typical limiting magnitudes reached are $g_J \sim21.5$, $r_F \sim21.0$, and
$i_N \sim20.3$, \ie, $\sim1^m - 1.5^m$ deeper than POSS-I.  The image
quality is improved relative to POSS-I, and is comparable to the Southern
photographic sky surveys.

The original survey plates are digitized at STScI, using modified PDS scanners \citep{las96}. The plates are scanned with
$15\mu$ ($1.0''$) pixels, in rasters of 23,040 square, giving $\sim1$
GB/plate, or $\sim3$ TB of pixel data total for the entire digital survey
(DPOSS).  Currently, astrometric solutions, provided by STScI, are good to $rms \sim0.5''$.

An extensive effort to process, calibrate, and catalog the scans, with the detection of all objects down to the survey limit, and star/galaxy classifications accurate to 90\% or better down to $\sim1^m$ above the detection limit, has been undertaken over the past several years. Object detection and photometry is performed by SKICAT, a software system developed for this purpose \citep{wei95a,wei95b,wei95c}, incorporating standard astronomical image processing packages, a commercial Sybase database, as well as a number of artificial intelligence and machine learning modules. Using this code, we measure $\sim60$ attributes per object on each plate. Nearly all plates at $|b|>10^{\circ}$ have been processed into catalogs. The catalogs are photometrically calibrated using extensive CCD sequences, with typical $rms$ magnitude errors of $0.25^m$ at $m_r=19.5$; see \citet{gal03a} for details of the calibration procedure. A fraction of this error is due to systematic offsets in the photometric zero points between plates. \citet{gal03a} show that the mean zero-point error is negligible, but has a $1\sigma$ scatter of $0.07^m$ in the $r$-band, which can produce significant plate-to-plate depth variance. Our solution to this potential problem is discussed in \S4. Star-galaxy separation is performed using a combination of FOCAS, neural networks, and decision trees, maintaining an accuracy of $>90\%$ at $m_r<19.5$ \citep{ode03}.

Each field in each band is processed individually. The three resulting catalogs are cross-matched to create a composite list of objects for the field. We require a detection in both the $J$ and $F$ bands so that we can measure the $g-r$ colors of our galaxies; this also reduces the likelihood that any object is a spurious detection. Areas on the plate containing saturated objects are masked. These areas often contain large numbers of falsely identified galaxies, as the plate processing software, tuned to find faint objects,  handles large, bright objects improperly.

In this paper we utilize a total of 237 plates with good calibration ($>1,000$ calibrating galaxies per plate from the CCD fields), at $b>30^{\circ}$, in the North Galactic Cap region. The distribution of DPOSS fields (as well as the detected clusters) can be seen in Figure 14. The total area coverage is 5800 square degrees. In addition, we have run our procedures on one field (475) from the SGP region of DPOSS, which is used to obtain additional spectroscopic follow-up (see \S5.3).

\section{The Detection Algorithm}

For this survey we use a modified version of the detection strategy first described in Paper I. We urge the reader to review Paper I for details of the adaptive kernel technique; we do not repeat the details here. In that earlier work, a color selection was first applied to the galaxy catalogs, after which we used an adaptive kernel technique \citep{sil86} to produce galaxy density maps, and a bootstrap technique was used to create significance maps. The FOCAS object detection algorithm \citep{jar81} was then used to detect density enhancements in the significance maps, which we identified as candidate clusters. We generated cluster catalogs for only two sky survey fields, for which we performed extensive follow-up imaging and spectroscopy. Since the publication of that work, we have gained a greater understanding of the plate-to-plate variance in DPOSS, significantly modified our photometric calibration techniques, and learned a great deal about our detection algorithms as a result of the additional data obtained. This has led us to significantly modify our procedures, while maintaining the underlying principles and methodology.

In Paper I, we utilized galaxies with magnitudes $m_r<20.0$. We have found that the random photometric errors, plate-to-plate zero-point offsets, and classification accuracy, can all reach unacceptable levels at that magnitude limit. Although some plates perform well to this limit, we found that imposing a slightly brighter magnitude limit, $m_r\le19.5$, and requiring a $g$ detection, produces a significantly more uniform galaxy catalog, without greatly sacrificing depth. Our intent with the current survey is to produce a catalog with good uniformity; therefore, we have elected to be rather conservative in the data utilized. Future work will use fainter objects to create a higher redshift cluster catalog which may not be suitable for large scale structure work, but which will provide a useful sample for cluster studies over a larger distance/time baseline. Additionally, in Paper I we imposed a liberal selection in $g-r$ color; we no longer apply this cut for the same reasons that we adopt the shallower magnitude limit. Because photometry in the $g$ and $r$ bands is completely independent, the use of a color cut effectively increases the pistoning due to zero-point offsets among plates by a factor $\sqrt{2}$. Finally, star/galaxy separation for DPOSS has been improved from the version used to produce the input data for Paper I, resulting in purer and more uniform galaxy samples being generated from the individual plate catalogs. The final catalog of objects now used in cluster detection therefore consists of all galaxies, with detections in both the $g$ and $r$ bands, having $15.0\le m_r\le19.5$. A typical galaxy catalog for a single plate contains $\sim 27,000$ galaxies over an area of $\sim34\Box^\circ$, resulting in a mean galaxy density of $0.79\times10^3$ galaxies per square degree. 

The resulting galaxy catalog is used as input to the adaptive kernel (AK) density mapping algorithm. As described in Paper I,  this technique uses a two-stage process to produce a density map. First, it produces an initial estimate of the galaxy density at each point in the map, which is then used to apply a smoothing kernel whose size changes as a function of the local density, with a smaller kernel at higher density. We refer the reader to Paper I for a more detailed description. As before, we generate our maps with $1'$ pixels. However, rather than the $900''$ kernel used earlier, we now use a significantly smaller kernel of $500''$ radius. Based on the simulations discussed in the next section, we found that this smaller kernel prevents over-smoothing the cores of higher-redshift ($z\sim0.3$) clusters, while avoiding fragmentation of most low redshift ($z\sim0.08$) clusters. The effect of varying the initial smoothing window is demonstrated in Figure 1. In this figure, we have placed four simulated clusters into a simulated background, representing the expected range of detectability in our survey. There are two clusters at low-$z$ (0.08), and two at high-$z$ (0.24), with one poor and one rich cluster at each redshift (100 and 333 total members, respectively). Of the 100 (333) total cluster galaxies, only 54 (186) are brighter than our magnitude limit ($m_r\le19.5$) at $z=0.08$, and 9 (17) at $z=0.24$. The corresponding richnesses ($N_{gals}$, see \S6.2) are 80 and 25. The initial smoothing window is varied from $300''$ to $800''$ in $100''$ steps. This figure clearly shows the segmentation of rich, low-$z$ clusters by small kernels, and the smoothing away of high-$z$ clusters by large kernels. 

Unlike Paper I, we do not perform bootstrap tests to produce significance maps. Cluster candidates are detected directly from the AK map of the actual data. In the current work, we use SExtractor \citep{ber96} to perform object detection (rather than FOCAS, as we did originally). We have found that SExtractor is both faster and more easily configured to meet our needs. For each plate, we find the set of SExtractor detection parameters (a pair of threshold level and minimum area) which produces an acceptable level of contamination by false clusters (10\%). The simulations used to determine these parameters are discussed in \S5.1. Once the set of optimal detection parameters for each plate is determined, further simulations are used to assess the catalog completeness as a function of redshift and richness, as discussed in \S5.2. We have opted to vary detection parameters between plates to maintain a constant level of contamination, even though the completeness may change. Due to the inhomogeneous nature of plate data, as well as large scale structure, setting a fixed threshold (in galaxy surface density) for all plates can lead to significant variance in both completeness {\em and} contamination between plates. By adjusting the parameters for each plate, we can minimize the variance of one of these quantities, and estimate the impact on the other. Because contamination is more easily quantified (as an integrated rate, independent of richness and redshift) as the fraction of detections that are false, we have chosen to adjust the parameters to maintain a fixed contamination rate. To allow large scale structure work, we provide a table (see Table 2) for each plate, including the RA/Dec boundaries used and the completeness function, as described in \S5.2.

Additionally, to assess the impact of photometric zero-point errors, and remove those cluster candidates whose detection is sensitive to these errors, we generate a set of ten galaxy catalogs for each plate with random zero-point offsets added to the $r$-band magnitude, drawn from the known photometric error distribution for DPOSS given in \citet{gal03a}. Like the original data, these catalogs are clipped at $15.0\le m_r\le19.5$, and an AK map generated for each one. Cluster detection is then performed using the same parameters as for the original map, and the resulting lists of cluster candidates are compared. Using the photometric redshifts (described in \S6) for each candidate from the original map, we match clusters within a 300kpc projected radius. In Table 1 we present the fraction of total candidates that appear in a given number of maps. Only those candidates which appear in the original catalog, and more than 7 of the 10 zero-point-error-added AK maps are kept in the final catalog; the number of simulated maps a cluster is found in (out of a maximum of ten) is provided in the final cluster catalog, and can be used to select increasingly certain subsamples. We use seven detections among the Monte-Carlo maps as our limit because the fraction of clusters detected in fewer maps is constant, whereas it drops steeply when more detections are required.

The requirement of $\ge7$ detections from the Monte-Carlo maps removes $<10\%$ of the cluster candidates. Because the detection of candidates that appear in fewer maps is very sensitive to small photometric errors, these candidates are likely to be either false, extremely poor, or at high redshift. For the latter two cases, our detection efficiency is very low (as described below), and such clusters should not be used in statistical studies. Thus, we expect that this requirement imposes no significant bias on the final catalog, and merely 
serves to reduce the catalog inhomogeneity at the limits of our detection procedure.

\section{Contamination, Completeness and Optimizing Detection}

The weakest aspect of optical imaging surveys for galaxy clusters is the lack of distance information. The two-dimensional projection of the galaxy distribution can produce many apparent overdensities which may be identified as clusters but are not actual physical associations. This issue has been discussed in some detail for the Abell catalog (see the overview in \S2), where rates of false clusters due to projection effects as high as 40\% and as low as 10\% have been claimed. Other types of surveys (\eg, X-ray, SZ) are not as strongly affected by projection, but suffer from other selection effects or observational difficulties. Therefore, redshifts are needed to validate the existence of a given cluster. 

The number of cluster candidates varies significantly with the detection
threshold \citep{kim02}. Obviously, this dependence is also related to
the number of false positive detections. As we try to achieve high
completeness we may suffer from high contamination. Our goal 
is to minimize $N_{false}/N_{cand}$, while maximizing $N_{cand}$, where 
$N_{false}$ is the number of false positives and $N_{cand}$ is the number 
of candidates for each field. We make use of a simulated background 
distribution in order to evaluate $N_{false}$, while $N_{cand}$ is derived 
from the real galaxy catalogs.  Because each plate suffers from slightly 
different systematic errors in both photometry and star/galaxy separation, our
procedure to optimize detection and evaluate completeness and contamination is done for each plate individually. This is preferable to selecting a single region of the sky, optimizing the algorithm for that region, evaluating the
contamination rate, and assuming that is true for the rest of
the sky. 

\subsection{The Rayleigh-Levy Distribution and Contamination}

Initially, we attempted to estimate contamination rates by producing random catalogs with the same mean density as our survey fields. This resulted in obvious underestimation of the contamination rate. As pointed out by \citet[hereafter PLOD]{pos02}, the basic sources for false
positives in cluster catalogs based only on photometric data are random fluctuations and superpositions of poor galaxy groups. For this survey, in which we effectively use a single photometric band for cluster detection, the susceptibility to false detections due to projection effects is high. Therefore, we adopt a methodology similar to that of PLOD, using the Rayleigh-Levy (RL) distribution to generate simulated galaxy positions. 
The RL random walk creates a linear ordering among the distributed
galaxies, implying that galaxies interact only with the ones in their
immediate vicinity, and each small set of neighbors interacts
independently of any other small set. The resulting galaxy
distribution, as shown by \citet{man75}, follows the same two-
and three-point correlation functions as described by \citet{pee75}.

For a given separation
$\theta$, the RL distribution gives the probability of finding a pair with a larger separation:

$$P(>\theta) = (\theta/\theta_0)^{-d} \hskip 0.3cm if \hskip 0.3cm 
\theta \geq \theta_0$$ and 
$$P(>\theta) = 1 \hskip 0.3cm if \hskip 0.3cm \theta < \theta_0$$

We ran a suite of RL simulations with different values for the parameters 
$\theta_0$ and $d$, and chose values for these such that the resulting pair distribution matches the real galaxy distribution for the DPOSS data. We use $\theta_0 = 500''$ and $d=0.6$. This value of $\theta_0$ also corresponds to the diameter of the smoothing kernel; angular correlations are thus removed at smaller scales. We find that the values of $\theta_0$ and $d$ do not vary significantly among plates, and therefore adopt the values above for all plates.

For each plate, We create
the simulated RL distribution by choosing a random position
within the plate limits that does not fall into a bad area,
and then selecting a separation from the above distribution. From this
new point in the frame we repeat the same procedure until seven galaxies
are selected. We then choose a new random location and select another
new point in the frame, repeating the same procedure until seven galaxies
are selected. This process goes on until we generate a number of positions 
equal to the total number of galaxies for that particular plate. The resulting
distributions are very insensitive to the number of galaxies
we choose before restarting. 

We also tested the full random walk distribution, where we start from one 
location, and continue relocating to new points with step sizes following the 
$\rm P(>\theta)$ distribution (as opposed to choosing a new random point, 
distributing a fixed number of points around it, and repeating). 
Again, the final distribution is very similar to the
restricted random walk; this  is because
the clustering scale we are interested in is well below the cutoff
in the two-point correlation function introduced by the restricted
random walk. By comparing the angular separation distribution
for these different sets we find them to be indistinguishable
up to scales of $2^{\circ}$. We therefore use the restricted RL in our simulations, as it is significantly less computationally intensive than the full random walk.

For each plate in the survey, we produce an RL distribution with the same number of objects as the galaxy catalog for that plate, with the same bad areas excised. A density map is generated from this distribution, and SExtractor is run with a range of threshold/minimum area pairs on this map, along with the map generated from the real galaxy catalog. We use detection thresholds ranging from 900 to 2200 galaxies per square degree, in increments of 20, and minimum areas of 30, 40, 50 and 100 square pixels (or arcmin). Initial tests using a broader range of parameters indicated that the optimal pair would always be found within these ranges. Once detection is completed on both the RL and real maps, we measure the contamination rate as $C=N_{false}/N_{cand}$, and select the parameter set for which $C$ is closest to 10\%, while maximizing $N_{cand}$; typically, this results in $8\%<C<12\%$. We performed tests using other values for the contamination rate; we found that completeness is not substantially improved by allowing higher $C$, while lower rates, such as $C=5\%$, produced more widely varying parameter sets due to small fluctuations in $C$. Importantly, enforcing a fixed contamination rate does not produce large plate-to-plate variations in the completeness; this can be seen by comparing the completeness functions for a large set of plates. Nevertheless, it would be inappropriate to use a mean completeness function for the entire survey. 

We then examined the optimal parameters derived for each plate. We found that minimum areas of 30 or 40 square arcminutes produced very similar results, in terms of the distribution of both the detection thresholds and final number of candidates per plate. We also examined the completeness functions (described below), and found that they were extremely similar for both values of the minimum area. Using a minimum area of 40 square arcminutes produced slightly lower variance in  the number of candidates per plate; we have therefore used this minimum detection area for all plates. Physically, this is sensible: we know that the depth and classification accuracy vary from plate to plate, which will affect the threshold in galaxies per square degree, but there is no reason to expect the sizes of clusters to vary from plate to plate. The larger minimum areas (50 and 100 square arcminutes) produce radically fewer cluster candidates.

We stress that the entire optimization 
procedure is performed separately for each field. The optimal parameters 
for a given field will not be appropriate for another. This 
is not due only to zero point differences or plate variations, but also 
to the large scale structure (LSS) in the universe \citep{bra00}.
Our final cluster catalog is designed to maintain a constant contamination 
level across the whole sky. One must also note that the RL distribution does 
not take into account inhomogeneities at large scales; this test is specifically designed to estimate the contamination rate due to chance projections of small groups or physically unassociated galaxies. Because it is impossible to know {\it a priori} which overdensities are true physical associations contributing to large scale structure, there is no way to use the actual catalogs to measure contamination rates. Only complete spectroscopic follow-up can determine the accuracy of our estimates; as described below in \S5, our own spectroscopy suggests low contamination rates, consistent with the 10\% threshold imposed. Additionally, the results of PLOD demonstrate that the RL estimates of false positive rates is verified by such spectroscopy.

An alternative method to estimate contamination rates is to shuffle the galaxy positions in a plate catalog, while maintaining the object magnitudes. This method  preserves some very large-scale correlations, while removing those at smaller scales. As shown in \citet{lop03}, this technique produces contamination rates (and the corresponding optimized detection parameters) which are very similar to those derived using the RL distribution.

\subsection{Completeness From Simulations}
Perhaps the most important feature of this cluster catalog is the detailed assessment of the selection function. In order to properly assess this, it is fundamental that each plate be treated individually. Because the detection parameters for each DPOSS field are determined separately, we 
assess the completeness as a function of redshift and richness individually
for each plate, and provide this information for each plate. When generating mock catalogs from simulations (or from other observations), one can then apply these selection functions to mimic potential systematic effects in our catalog.

To assess the selection function for each plate, we use simulated galaxy clusters which are placed in the background 
field derived from the respective plate. We then use these galaxy catalogs consisting of a background plus 
artificial clusters 
 to generate density maps on which we run SExtractor. When 
running SExtractor, we use the same detection parameters obtained
in the previous section, which are also used for the real cluster detections.

Artificial clusters are typically defined using a Schecter luminosity function
with its choice of $\alpha$ and $M^*$; a surface density profile;
a given cluster composition; a maximum radius and a core radius. 
\citet{lob00} tested the variation of the completeness rate as a 
function of the spatial profile slope, while PLOD also tested different
cluster compositions. As expected, steeper profiles provide the highest completeness rates. We
discuss the effects of varying different cluster parameters at the end
of this section. 

An important issue is the background into which we insert the artificial 
clusters. While the RL distribution is well suited
to estimate the contamination rate and optimize the detection parameters, it 
only represents an ideal case, being unable to reproduce all
the inhomogeneities due to large scale structure. As we discuss
below, the RL distribution overestimates the completeness rate. Thus, for each plate, we use the actual galaxy catalog from that plate to provide the background field in which we insert the artificial clusters 

The simulated clusters follow a Schecter luminosity function with
parameters given by \citet{pao01} ($\alpha=-1.1$ and 
$M^*_r=-21.53$). The constituent galaxies have $-23.4\le M_r \le-16.4$, such that 
our observed magnitude range is fully covered at all redshifts of interest. 
The surface profile adopted is a power law in radius of the form $r^\beta$,
where $\beta = -1.3$. This is in the middle of the observed range, $-1.6 \le \beta \le -1.0$,
as discussed in \citet{tys95} and \citet{squ96}. The galaxies 
are placed within a maximum radius of $r_{max} = 1.5 h^{-1}$ Mpc with
a core radius of $r_{core} = 0.15h^{-1}$ Mpc.
The clusters are composed of 60\%  elliptical
galaxies and 40\% Sbc galaxies, with SEDs taken from \citet{col80}, convolved with the DPOSS $r$ filter. Each galaxy has the appropriate $k$-correction
applied, as well as a random photometric error from \citet{gal03a} added.

We proceed as follows:

\begin{enumerate}
\item We generate clusters with six different richnesses, which are shifted to seven redshifts, with the catalogs trimmed at $15.0\le m_r\le19.5$. The richness classes are $N_{gals}=[15,25,35,55,80,120]$ galaxies, while the redshifts adopted are
$z = [0.08,0.12,0.16,0.20,0.24,0.28,0.32]$.

\item For each of the 42 richness/redshift combinations we generate 5 simulated clusters, which are placed at random positions in the real background galaxy distribution, avoiding both excised areas and cluster candidates detected in the real fields. This is repeated ten times, resulting in ten catalogs containing a total of 50 simulated clusters for each richness/redshift combination. Thus, there are 420 total simulated catalogs generated for each plate (6 richnesses $\times$ 7 redshifts $\times$ 10 realizations). An example is shown in Figure 2, where we plot Plate 389, with the bad areas marked, and five simulated clusters at $z=0.16$ and $N_{gals}=80$ inserted. 

\item  We then use the AK to produce density maps for each of these 420 
background plus cluster galaxy catalogs. This is performed individually for each plate, and SExtractor run with the optimal parameters determined from the contamination tests.

\item Next, we compare the candidate positions in these maps with the initial input positions. We use a matching radius of $400h^{-1}$ kpc, which is small compared to the typical size of a cluster.

\end{enumerate}

\subsubsection {RL {\it vs.} Real Backgrounds}

For one DPOSS field (389) we repeated the entire procedure
twice, first using the real background and then the RL
background distribution. The purpose of this test is to check
how well the RL distribution reflects the real background. 
We show the results of this test in Figure 3, where the completeness rates obtained using the RL distribution are the dashed lines, while the solid lines use the real background. Each panel shows a different richness class, with richness increasing from bottom to top. We see that the dashed lines (RL
 background)
are always higher than the solid lines, for all richness classes and
at all redshifts. At $z$ = 0.16 (the median redshift of the DPOSS sample) and 
$N_{gals}$ = 35 (close to the median richness value), the recovery rate
is 94\% using the RL background, but only 72\% using the real background. This suggests that using the RL distribution likely overestimates the completeness.
An interesting comparison can be made with Figure 11. For each richness class
shown there, the median redshift  is similar to the redshift where the completeness estimate for that subsample starts to decrease significantly.

We note that \citet{kim02} modified their detection limits
after comparing the recovery rate of artificial clusters using  uniform and
 real backgrounds. Their goal was to improve the completeness rate; however, this necessarily has an adverse impact on the contamination. We do not apply such corrections, as our goal is to maintain a constant contamination rate.

\subsubsection{Dependence on Cluster Model}
We also wish to test the variation of completeness
with the cluster composition, surface density profile slope, luminosity function
slope, and maximum and core radii. We use a single plate, varying the 
properties of the simulated clusters and repeating our completeness tests. Figure 5 shows the dependence of the completeness on the cluster composition
and spatial profile slope ($\beta$). Each row represents a different value for
the slope ($\beta$ = -1.0, -1.3, -1.6, from bottom to top), 
while each column shows the results for different compositions (from right to left: 100\% E, 60\% E + 40\% Sbc, and 20\% E + 80\% Sbc). 
The improvement when we adopt steeper profiles is evident; as expected, tighter cores are easier to detect. There is little if any correlation  with the cluster composition. This is expected
as the difference in the $k$-corrections used (for E and Sbc galaxies) 
is only $0.2^m$ at the highest redshift of the simulations ($z=0.32$).
For instance, for $N_{gals}=35$ and $z=0.16$, using
60\% E + 40\% Sbc, the recovery rates for $\beta$ = (-1.0,-1.3,-1.6) are (62\%,72\%,80\%), respectively.
However, when we fix $\beta=-1.3$, $N_{gals}=35$ and $z=0.16$, the recovery rates are (74\%,72\%,70\%) for cluster compositions of (100\% E, 60\% E + 40\% Sbc, 20\% E + 80\% Sbc).

The effects of changing the luminosity function slope, core radius and
maximum radius are shown in Figure 6. Each row shows the results for one of 
these 3 parameters.  For the richest clusters, steepening the LF slope increases completeness, while poor clusters show an opposite trend, albeit with large scatter. The variation
with core radius shows a clear trend to higher completeness for
smaller values of this parameter; this is commensurate with the results of the radial profile test. Finally, tests with the cutoff radius show the worst
recovery rate for $r_{max} = 1.0h^{-1}$ Mpc, while the results are about
the same for the other two values tested.

We conclude that our canonical cluster model ($\alpha=-1.1, M^*_r=-21.53, \beta=-1.3, r_{max} = 1.5 h^{-1}$ Mpc, $r_{core} = 0.15h^{-1}$ Mpc, 60\%E + 40\%S) provides a realistic estimate of the catalog completeness. For each candidate, we provide a table of the completeness as a function of richness and redshift. An example is shown in Table 2 and Figure 4; a similar table is provided online (at \texttt{http://dposs.caltech.edu/dataproducts/}) for each plate used in this survey. Additionally, a list of the areas excised due to bright stars, airplanes, etc. is provided for each plate at the same location.

\subsection{Spectroscopic Confirmation}

To assess the validity of the above results, as well as to calibrate our photometric redshift estimator (see the next section and Paper I) and examine the redshift distribution of our clusters, we undertook a complete spectroscopic survey of candidates in two DPOSS fields: 447 ($14^{h} 30^{m}, +30^{\circ}$) and 475 ($01^{h}, +25^{\circ}$). These fields were chosen because they are at relatively high galactic latitude ($+67^{\circ}$ and $-40^{\circ}$) where the effects of dust are expected to be small, and because scans in all three bands were available when this project was started. However, these maps were generated, and cluster detection performed, using the methodology discussed in Paper I, which (as discussed earlier) differs significantly from the final technique used here.  The original list of cluster candidates in these two fields was drawn from the individually calibrated plates, using galaxies to $m_r=20$.  Therefore, the sample with follow-up spectroscopy is only a subset of the final cluster catalog in these areas, and also contains targets which are no longer candidates. We remind the reader that the catalog for Field 475, which is in the SGP, is not a part of the NGP, and thus not in the area covered by the catalog presented in this paper.

\subsubsection{Observations}
Spectra of the cluster candidates were obtained between April 1996 and April 2000 with the COSMIC instrument \citep{kel98} in re-imaged mode, mounted at the prime focus of the Hale 5m Telescope at Palomar Observatory. Targets for each mask were chosen from the DPOSS catalog for the appropriate field, selecting only galaxies with $m_r\!\leq\!20$ to maintain reasonable exposure times. No color selection was applied; this was done to maximize the number of possible objects which could have slits placed on them.
Multi-slit masks, made of photographic film negatives, with $1.5''$ slit widths,  were mounted at the focal plane. The re-imaged pixel scale is $0.399''$/pixel; the slits therefore correspond to $\sim\!3.75$ pixels. The available field for spectroscopy is theoretically $8'\!\times\!12'$, with the long dimension corresponding to the spatial axis. The field is further limited by the lack of an accurate distortion map in the slitmask production software; this requires that slits be placed within $\sim\! 3'$ of the central axis in the direction perpendicular to the slits. Nevertheless, the $8'\times3'$ field of the masks is larger than the core radius of clusters at all but the lowest redshifts in our sample.

Slitmasks were designed using the $cosmicslitmask$ program, written by M. Pahre. This software simply takes an input list of coordinates, and allows the user to interactively select objects for slit assignment. There is no automated slit optimization algorithm, so objects were selected visually for each mask, to maximize the number of targets observed per mask. This often resulted in densely packed slitmasks, with as many as 45 slits on a single mask, some as short as $10''$. 

Spectra were taken using a grism with 300 l/mm, blazed at 5500\AA, which produced a dispersion of 3.03\AA~ pixel$^{-1}$. The spectra typically cover a wavelength range $\sim\!3000$\AA~ to $\sim\!9000$\AA, with a central wavelength of $\sim\!6000$\AA. Since the dispersing element is a grism and cannot be tilted, the wavelength range is fixed and varies only as a result of the vertical displacement of the slits from the mask center. Therefore, nearly all spectra are observed from blueward of the 4000\AA~ break (for $z=0$), through the redshifted Na lines (5890\AA~ rest frame, 7950\AA~ at our highest expected $z=0.35$). The COSMIC CCD has a gain of $3.1e^-/DN$ and a relatively high read noise of $13e^-$.  Exposure times varied greatly from night to night, depending largely on the seeing (which ranged from $0.8''$ to as high as $2.0''$) and spectrograph focus changes (due to temperature fluctuations). Exposures ranged from a single 1800s exposure up to two 3600s exposures per slitmask, with most consisting of two 1800s exposures.

\subsubsection{Data Reduction}

All data reduction was performed using the IRAF package. The object spectra were overscan subtracted, and flattened using spectra of a halogen lamp reflected off the dome interior. Cosmic rays were removed using the {\it szap} task (written by M. Dickinson), and night sky emission removed by subtraction in the spatial direction on the two-dimensional images, using the {\it background} task to perform median filtering and sigma clipping. Because the slits were typically very narrow, we used only a second order fit along the spatial direction. After sky subtraction, each spectrum was traced and optimally extracted in the individual exposures. The extracted spectra were wavelength calibrated using spectra of an arc lamp, taken immediately before or after the object spectra, to minimize the effects of instrument flexure. The individual extracted, wavelength-calibrated spectra were then combined to produce the final spectrum for each object, used to measure the redshifts. All redshifts were measured by RRG using the IRAF task {\it redshift}, written by T. Small. This program allows the user to visually mark specific spectroscopic features, and input their rest wavelengths. The object's redshift is then calculated, and the positions of other common absorption and emission lines overplotted. In this way, the user can check the redshift assignments. Such visual measurement was necessitated by the poor $S/N$ of the majority of our spectra; automated techniques (such as cross-correlations) were strongly affected by residuals from poor sky subtraction. To enable completion of the survey in a reasonable time,  a trade-off in exposure times {\it vs.} number of masks observed was made, since we were only interested in measuring redshifts and not any specific galaxy properties (such as velocity dispersions or line strengths). Comparison of redshifts obtained by different reducers and at different times suggests the redshifts are accurate to $\sim\!\Delta z\!=\!0.002$, or approximately 600 km s$^{-1}$.

In addition, over the course of the survey, our list of candidate galaxy clusters changed as the photometric calibration and sample selection improved. This resulted in a significant number of targets being observed which are not included in the final cluster sample. This extraneous data is actually useful in assessing the rate at which we miss real clusters and is discussed later in this section. A total of 3249 individual spectra were visually inspected; of these, 1655 (51\%) were identified as galaxies, and 326 (10\%) as stars. Thus, a total of 1981 spectra were identifiable, a 61\% success rate. Of the galaxies, 1245 (75\%) had securely measured redshifts; the remainder had insufficient $S/N$. We also note that the 10\% stellar contamination is consistent with the results of \citet{ode03}, and should be taken as an upper limit on misclassification, as extended sources (\ie, galaxies) tend to have lower $S/N$ spectra, making identification more difficult. This large spectroscopic sample constitutes a significant survey in its own right, with other applications in addition to those discussed here.

In Field 447, there are 64 cluster candidates. Of these, only 24 have usable spectroscopy, with all 24 showing evidence for clusters. There are an additional 4 masks targeted at areas with no current candidate; these nevertheless show spectroscopic evidence for physical galaxy associations. However, they all have $z\ge0.22$ and $N_{gals}<30$, a regime where our completeness is typically only $\sim30\%$. There are unfortunately 40 current candidates in this field without spectroscopy.

The situation for Field 475 is much better, with 55 candidates in our current sample, of which 37 have corresponding slitmasks. These all show evidence for physical clustering (as discussed below). There are 16 masks not associated with any new candidates. Of these sixteen, four are near the edge of the plate or the area used for the densitometry spots. There remain twelve masks which show evidence for real galaxy associations. Nevertheless, these twelve areas appear to be only moderate overdensities; even lowering our detection threshold to allow 30\% contamination recovers only three of these. Running our richness estimator on these areas, using the spectroscopic redshifts, shows that indeed these possible clusters all have $N_{gals}<30$. Additionally, four of these 12 masks have clusters at $z\ge0.22$. Finally, eighteen current candidates have no spectroscopy; these are all of similar overdensity to those candidates with spectra, and therefore likely to be real.

The results of the spectroscopic survey are summarized in Table 3. We give the total number of candidates, the number (and fraction) with and without spectroscopy, and the number of extraneous masks. Details for each candidate with spectroscopic observations are provided in Tables 4 (for Field 447) and 5 (for Field 475). Column (1) provides the candidate name, columns (2) and (3) provide the spectroscopic redshift and the number of galaxies at that redshift (if spectra were obtained), and column (4) provides comments, if any. 

The above results are shown in Figure 7, for Field 447 (left panel) and 475 (right panel). We show the AK galaxy density maps, with blue circles marking the locations of current candidate clusters. These clearly correspond to areas with the highest galaxy density. Green circles show the locations of slitmasks associated with current candidates, while red circles mark locations where spectra were taken but there is no current candidate.

We show example spectroscopic redshift histograms for some cluster candidates in Figure 8. From these redshift distributions, we can clearly see strong clustering in redshift space, providing evidence for true galaxy clusters in most of the slitmasks. The median redshift of our spectroscopically confirmed cluster candidates is $z_{med}\!\sim\!0.2$. This value is in keeping with the photometric redshifts presented in the following section, and with our earlier estimates of the depth of our galaxy catalogs.

\subsubsection{Are Redshift Peaks Real Clusters?}
Although the peaks in the redshift distributions provide evidence for clustering, we would like to quantify the likelihood that a given peak is a real structure. Certainly, large peaks (as in candidate NSC005559+262442, with 24 galaxies) are undoubtedly clusters. However, many of the masks show peaks with only 3 or 4 members. Individual galaxies are assigned as members of a peak if they fall within $\Delta z=0.005$ ($\pm1500$ km s$^{-1}$) of the peak center. Increasing this range does not change the results significantly, since most redshift peaks are well isolated.

A rough estimate of the likelihood for such small peaks being real can be made following the argument of \citet{zar97}. If the underlying redshift distribution of galaxies were smooth, the formal probability of a pair of galaxies within $\Delta z=0.005$ would be negligible, much less for three or more galaxies. However, because galaxies are correlated, the likelihood of pairs, triplets, etc., is increased. If we assume that {\it all} slitmasks where the richest clump has only three or four members correspond to spurious detections (there are 18 such masks out of 95 total masks), the probability of finding a single false peak is $18/95$, or $\sim\! 20\%$. We might then expect a comparable fraction of our histograms to show a second redshift peak with similarly few members, but there are only eight ($\sim\! 8\%$). In fact, some of the distributions with multiple peaks may be projections of multiple poorer clusters which we detect as a single candidate. Additionally, there is only one mask with a secondary peak containing five members.  An additional argument in favor of the reality of three-or-more galaxy clumps is that if peaks with three members were often spurious, we expect to find many more such peaks than those containing four galaxies. However, we find an almost identical number of peaks with three and four galaxies, supporting the idea that even three-galaxy peaks are often real. Therefore, we conservatively identify all peaks with four or more members as real clusters, and peaks with only three members as tentative. The number of clumps with two galaxies is much higher (nearly every mask contains one); we therefore ignore those completely. Furthermore, while Zaritsky et al. follow similar arguments for only two galaxies, their spectroscopy covers a significantly larger magnitude range ($m_r\!<\!22$ compared to our $m_r\!<\!20$), thereby greatly increasing their likely contamination rate. Conservatively, we estimate that at least 80\% of our robust candidates are true clusters; this is roughly consistent with our attempt to generate catalogs with $\sim10\%$ contamination.

\section{Photometric Redshifts and Richnesses}

Two of the most fundamental properties of galaxy clusters are their distance (or redshift), and mass. Because the latter is difficult to measure accurately, and impossible from our photometric data alone, we instead measure the richness. The details of our redshift and richness estimation techniques  are given below. It is important to note that both are measured in a completely model independent way; this is possible because, these 
properties are measured {\it after} detection is completed, and are not an inherent part of the procedure. We note that this is inherently different from many other techniques; for instance, the Matched Filter simultaneously performs detection and redshift/richness estimation, while others, such as the Cut \& Enhance method \citep{got02} or the RCS \citep{gla00} output the redshift, based on galaxy colors used as part of the detection procedure.
 
\subsection{Redshift Estimates}
Paper I presented an extremely simple yet effective photometric redshift estimator for DPOSS cluster candidates, based on the assumptions that each candidate is a single cluster, at one redshift, and the cluster galaxy population is dominated by early-type galaxies. The estimator is an empirical relation between spectroscopic redshift and the median $g\!-\!r$ color and mean $r$ magnitude of the background-corrected galaxy population for each candidate. We use both colors and magnitudes as we find both to be equally well correlated with the spectroscopic redshift. We count the number of galaxies as a function of color, $N_{g-r}$, and the number as a function of $r$ magnitude, $N_r$, inside a radius of 1 Mpc ($0.67\!\times\!R_{Abell}$). The background galaxy color and magnitude distributions ($N_{bg,g-r}$ and $N_{bg,r}$) are determined independently for each plate, scaled to the appropriate area, and subtracted from the color and magnitude distributions of each candidate cluster. The median $g\!-\!r$ color and mean $r$ magnitude of the remaining galaxies is then calculated. This differs slightly from the methodology in Paper I, where universal background distributions, taken from a large contiguous area, were utilized. In practice, this must be an iterative procedure, because we do not initially know the redshift, and therefore the Abell radius, for our cluster candidates. We therefore start with an initial radius corresponding to 1 Mpc at $z=0.05$, within which we measure the above quantities and derive the initial redshift estimate. The Abell radius is adjusted using this new redshift estimate, and the procedure repeated until the redshift converges.

Spectroscopic redshifts for the clusters used to derive our empirical photometric redshift estimator were taken from \citet[hereafter SR99]{str99}. We do not use our own spectroscopic data, as they cover only two plates, and we do not want to bias our estimator. The SR99 data are culled from diverse sources in the literature, and are of highly variable quality. A majority of the cluster redshifts, especially those at $z\!>\!0.1$, are based on very few (one or two) individual galaxy redshifts, and may therefore be incorrect. Nevertheless, this catalog is the largest, somewhat homogenized resource currently available. The larger area utilized in this paper results in a final training sample comprising 369 clusters (compared to only 46 in Paper I), with a median redshift of $z_{med}=0.138$. Because we now have many more clusters, and their redshift distribution is heavily weighted with low-redshift clusters, we bin the colors and magnitudes into spectroscopic redshift bins. We divided the range $0.02\le z_{spec}\le 0.35$ into ten bins, and calculate the median $z_{spec}$, $(g-r)_{med}$, and $r_{mean}$ for each bin. These binned values are used to derive an empirical relation between redshift, median $g\!-\!r$ color, and mean $r$ magnitude using a bivariate least-squares fit:
\begin{equation}
{z_{phot} = 0.5694\times(g-r)_{med} - 0.00215\times r_{mean} - 0.057243 }
\end{equation}

The upper panel of Figure 9 shows the photometric redshift against the spectroscopically measured redshift for these 369 Abell clusters, with the lower panel showing the residual. Performing the complete iterative procedure, where we begin without assuming the spectroscopic redshift to define a starting radius, yields a $Q_{sigma}$ of $(z_{spect} - z_{phot})/(1+z_{spec}) = \mathrm{\Delta}z=0.033$.  

In addition, we tested the effect of varying the starting cluster radius from $5'$ (corresponding to $\sim1.0$Mpc at $z=0.2$) to $15'$ ($\sim1.0$Mpc at $z=0.05$). In principle this could lead to different final photometric redshifts if the convergence procedure is unstable. The results are shown in Figure 10; the $Q_{sigma}$ for $\mathrm{\Delta}z_{phot}=0.004$. There are a very few outliers, which demonstrates the robustness of our simple technique. This also suggests that estimating $z_{phot}$ with two different initial radii and selecting those with varying results can be used to test for clusters where the estimated redshift is questionable. The starting redshift also does not affect the accuracy of the estimator.

The photometric redshift estimates for the candidate clusters are provided in the fourth column of Table 6. Those clusters where the photometric redshifts using the two starting redshifts disagree by more than $\sqrt{2}$ times the redshift error are marked with a colon; in cases where the estimator failed entirely, we set $z=0.0$. The mean redshift for our sample is $z_{phot,med}=0.1579$, which is comparable to our original estimate of $<\!z\!>\sim\!0.15$, based on the magnitude range covered. Our highest redshift clusters are at $z=0.3$, with many at $z\!>\!0.2$. The redshift distribution of our robust sample is shown in Figure 11. The entire sample is shown in the top panel, with subsamples of varying richnesses in the lower bins. The median redshift for each sample is marked. As expected, poor clusters are found only at low redshift, while richer clusters can be seen to larger distances.

\subsection {Richness}
 
A fundamental physical property of galaxy clusters is their mass. Mass can be measured using X-ray data, velocity dispersions, lensing (strong and weak); all of these methods require data which is difficult to obtain for extremely large cluster samples, and often impossible for low-mass systems. One expects that the number of galaxies in a cluster (the cluster richness) should be correlated with the total cluster mass. Unfortunately, cluster richnesses are notoriously difficult to measure, and there are a variety of different estimators. Abell's richness, for instance, is extremely poorly correlated with mass; other estimators fare somewhat better \citep{yee99}. Because our catalog covers a large fraction of the sky, we can measure optical richnesses for many clusters which have spectroscopic data, as well as prepare specific subsamples for future spectroscopic study.   

Given a redshift estimate for each cluster, we can measure a richness that samples the same absolute magnitude range. We compute the richness in a fixed absolute magnitude interval, $M^*_r-1\!<\!M\!<\!M^*_r+2$, using the $r$-band data, and assuming $M^{*}_{r}=-21.53$, taken from the cluster luminosity functions derived by \citep{pao01}.  Galaxies are counted within a radius of 1$h^{-1}$ Mpc. For clusters at low redshift ($z\!<\!0.17$), we measure the richness by directly summing the background-corrected number of galaxies in the appropriate magnitude interval. Only at these low redshifts is this entire magnitude range contained within our data. For more distant clusters ($z\!>\!0.17$), we directly sum the galaxies with $M^*_r-1\!<\!M\!<\!M_{20}$, where $M_{20}$ is the absolute magnitude limit corresponding to our catalog limit $m_r=20.0$. We then calculate a correction factor $\gamma$ for the richness, defined as

\begin{equation}
\gamma = \frac{\displaystyle{\int_{M^*_r-1}^{M^*_r+2}}\Phi(M)dM}{\displaystyle{\int_{M^*_r-1}^{M_{20}}}\Phi(M)dM}
\end{equation}
We set the exponent in the luminosity function to $\alpha\!=\!-1.1$, as we do for our simulated clusters; the correction factor is slightly larger for lower $\alpha$ (by $\sim\!5-10\%$ for $\alpha\!=\!-0.87$). Typical values of $\gamma$ are 1.125 for $z\!=\!0.2$ and 1.50 at $z\!=\!0.3$. 

We use a background determined individually for each plate. We simply take the distribution of galaxies as a function of magnitude for the entire plate, and subtract this (scaled to the appropriate area) from the same distribution for each cluster. This ensures that any systematic errors in an individual plate (calibration, classification) are maintained when performing the background
subtraction. Our methodology lies  between completely local background estimation (such as taking an annulus near the cluster), and completely global estimates (which would use the entire survey area). Tests using median filtering to remove overdense regions from the background measurement show that this method does not change the final richnesses appreciably; the mean richness is increased by $\Delta N_{gals}=3$) if this alternative method is used.

The final richness ($N_{gals}$) for each cluster is listed in the fifth column  of Table 6. We show the richness distribution of our robust cluster candidates in Figure 12, for the whole sample and in different redshift bins. At high $z$, where our completeness drops, we find only richer clusters, while at low $z$, where the volume is smaller, we find mostly poor clusters. The median richness for our clusters is $N_{gals,med}=31$, which corresponds to the poorest end of Abell's richness class 0. We therefore expect that our catalog contains many very poor clusters (or even groups). 

\subsubsection{Richness Errors}
Errors in our photometric redshift estimator will introduce an error in the measured richness for each cluster. We investigated the magnitude of this effect by calculating richnesses for the Abell clusters with spectroscopic redshifts. We calculate the richness using both the spectroscopic redshift and our photometric redshift. The results are shown in the upper left panel of Figure 13. The two estimates are very well correlated, with no systematic offset, and larger scatter at lower richness, where the redshift estimates are likely to be worse. 

We also tested various cluster radii for measuring richness. The upper right panel of Figure 13 shows the ratio of richnesses using 1.5Mpc and 1Mpc radii, versus the 1.5Mpc radius richness. We see that the ratio is approximately constant regardless of richness, and that the ratio is $\sim1.3$, less than the ratio of the areas. This is clearly because the outer parts of the clusters have low density. We therefore use the 1Mpc radius richnesses, as the background subtraction over a larger area will introduce more noise.

There are many other potential sources of error, both random and systematic. A future paper \citep{gal03b} will address many of these issues in detail, including the choice of background area, cosmology, $k$-corrections, the radius used, etc. We will measure cluster richnesses using a variety of techniques; this will allow detailed tests to find which methodology provides the best surrogate for a mass measurement. Nevertheless, we are confident that our {\it ad hoc} estimator provides a useful measurement. We have compared our richness estimator to the $\Lambda_{cl}$ measurement of \citet{kim01} for clusters detected by both surveys. The results are shown in Figure 14, where solid circles are clusters where the redshift estimates agree (within errors), while open circles are clusters where they do not. The two richness 
measures are extremely well correlated, suggesting that they are both measuring similar properties of the clusters. The agreement is quite remarkable, as they are based on different surveys, and measure different properties. One method ($N_{gals}$) simply counts galaxies, while the other considers the luminosity of the galaxies ($\Lambda_{cl}$).

\subsubsection{Comparison to Abell}
Finally, we attempted to compare our richness estimate to those of Abell. The results are shown in the lower left panel of Figure 13, with each comparison cluster as a small dot, and the boxes representing binned data. Although a correlation is evident, the scatter is large; significantly larger than the errors due to our measure alone. Abell's richness estimate, counting from the third brightest galaxy to two magnitudes fainter, suffers from many drawbacks. First, his redshift estimates are often incorrect, causing errors in the radius used to count galaxies. The use of an apparent magnitude, $m_3$, to set the magnitude limits means that different clusters have their richnesses measured in different absolute magnitude ranges, and introduces random errors due to bright foreground galaxies. This effect is shown in the bottom right panel of Figure 13, where we plot Abell's richness against our estimate of the cluster richness measure using a version of his technique (detailed in \citealt{kim01}). A similarly poor, but extant, correlation can be seen. The difficulties with Abell's richness measures are well documented, as discussed in \S1. 

\section{The Cluster Catalog}
The first installment of the Northern Sky Optical Cluster Survey catalog presented here covers $5834\Box^{\circ}$, and contains 8,155 cluster candidates, yielding 1.4 clusters per square degree. In comparison, Abell's 1958 survey covered a much larger area, yet contains only 2,712 clusters. Our catalogue thus represents an increase of roughly one order of magnitude in cluster counts. More importantly, these clusters have been selected using an automated, objective algorithm, with extremely well characterized contamination rates and selection functions. Each cluster also has a consistently measured richness and photometric redshift.  Figure 15 shows the sky distribution of all our cluster candidates. The median redshift of our sample is $z_{med}=0.1579$, with a median richness of $N_{gals,med}=31$. Thus, our sample is both somewhat deeper than Abell, and extends to significantly poorer systems. As mentioned earlier, the redshift distribution is shown in Figure 11, and the richness distribution in Figure 12. DPOSS $F$-plate images (500'' diameter) of four new, rich clusters are shown in Figure 16. Two were not previously known, and two (NSC172013+264028 = RXC J1720.1+2637 and  NSC122906+473720 = RXC J1229.0+4737) were detected only as X-ray clusters.

We provide the complete cluster catalog for our survey area in Table 6, as well as at \\ \texttt{http://dposs.caltech.edu/dataproducts/}, sorted by increasing RA. 

The table provides:
\begin{enumerate}
\item The cluster name. The naming convention is NSChhmmss+ddmmss. Coordinates are J2000.
\item The cluster RA and 
\item The cluster Dec, in decimal coordinates, also J2000.
\item The photometric redshift. If this is zero, the estimator failed to converge.
\item The measured richness. If $z_{phot} > 0.17$, the correction factor $\gamma$, as described in \S6.2, is applied. 
\item The number of simulated maps in which the cluster was detected. Only clusters detected in seven or more maps are included here.
\item The plate number from which the cluster is drawn.
\end{enumerate}

\subsection{Consistency with Previous Surveys}
Although our data and methodology is different from those of prior surveys, we expect that optically-based cluster searches will find similar objects. Thus, we may ask if our results are consistent with other recent surveys. 

The obvious survey to compare with is that of Abell. In the area covered in this paper, there are 1090 Abell clusters. Of these, $\sim93\%$ are recovered by us; an exact number is difficult to determine due to centroiding errors, varying matching radii, unknown redshifts, proximity to excised areas, and other effects; some of these will be treated in a full comparison when the second (and final) portion of our catalog is completed. Nevertheless, the high recovery rate suggests that the Abell catalog consists mostly of real clusters (as argued by \citet{mil99}), but is incomplete, as our sample contains eight times more clusters. Unfortunately, it is difficult to assess the Abell catalog incompleteness in the regime where it is used for statistical studies ($N_{Abell}>50$, $z<0.2$) because of the poor correlation between Abell's richness and our $N_{gals}$. If we assume $N_{gals}$ and $N_{Abell}$ are equivalent, then the two catalogs contain very similar numbers of clusters in the aforementioned range. However, a small shift in the limits used for comparison (for instance, setting $N_{gals}\ge45$) would lead one to conclude that the statistical Abell catalog is $\sim35\%$ incomplete. These discrepant results are consistent with the difficulties encountered by other authors in assessing the completeness of the Abell catalog. 

An excellent alternative comparison sample is that of PLOD, who discuss the surface density of clusters as a function of redshift. They find 1.19 clusters per square degree for $z\le0.2$ and $\Lambda_{cl}\ge40$. In our 5800 square degree sample, we therefore expect 6902 clusters for the same range of parameters. Unfortunately, we do not measure $\Lambda_{cl}$, but PLOD provide a conversion to Abell richness, $\Lambda_{cl}=1.24\times R_{Abell}$; their $\Lambda_{cl}$ cut thus corresponds to $R_{Abell}\gsim30$. At these low richnesses, the Abell richness corresponds roughly to our $N_{gals}$ (see Figure 13). We therefore apply these cuts ($z_{phot}\le0.2$, $N_{gals}\ge30$) to our catalog, resulting in a sample of 2888 clusters. We then apply our selection function (shown in Figure 3); at the higher end of the redshift range ($z\sim0.15$) our detection efficiency
is quite low for the most common, poorer clusters. The estimated actual number of clusters expected in our survey area, after applying this correction, is 6461 clusters, compared to 6902 predicted by the constant comoving space density given by PLOD. This is only a 6\% difference, which is remarkable given that both surveys use different techniques, and have to apply estimated selection functions to arrive at actual space densities.

This excellent agreement suggests that both our survey and the smaller but deeper survey of PLOD detect similar objects in the overlapping redshift range. In addition, the selection functions presented by the two surveys provide realistic estimates of their completeness.

We note also that PLOD find a space density of Abell-like clusters that is a factor of 1.5 higher than in the Abell catalog. This result relies on a conversion of $\Lambda_{cl}$ to $N_{Abell}$, which, like our own conversion, is somewhat uncertain. Nevertheless, the agreement in density between our survey and PLOD does suggest significant incompleteness in the Abell catalog. Because the number of detected clusters is strongly increasing at the poor end of Abell's $R=1$ bin, these results are very sensitive to small changes in the richness cuts applied.

A preliminary comparison to the NORAS X-ray cluster survey \citep{boh00} shown in \citet{gal01} demonstrates that nearly 90\% of NORAS clusters at $z<0.2$ are detected in DPOSS. In contrast, the DPOSS cluster catalog contains over an order of magnitude more clusters than NORAS; in most cases these are poor systems. To understand the biases in these catalogs, we are undertaking a full comparison in which the same properties ($f_X$, $N_{gals}$, $z_{phot}$) are consistently measured for both NORAS and DPOSS clusters.

\section{Discussion and Future Directions}

We have presented the community a large, new cluster catalog, which meets the original objectives outlined at the beginning of this paper. This catalog includes accurate photometric redshifts and richnesses for a sample of 8,155 cluster candidates. Additionally, we provide detailed selection functions in both redshift and richness, on an individual plate basis, which can be used in the generation of mock catalogs for testing cosmological models, measuring the
correlation function, and other large scale structure studies.

Certainly, this will be superseded in the future by deeper surveys, with more accurate photometry in more bandpasses, and with superior spectroscopic or photometric redshift information. Our catalog is still limited by the use of the projected galaxy distribution to detect clusters, and the poor photometric accuracy in comparison to modern digital surveys, such as the SDSS.

In the future, we will publish an additional catalog, covering the Southern Galactic Cap region, and including the less well calibrated plates from the NGP. The final Northern Sky Cluster catalog will then cover over 10,000 square degrees.

The scientific uses for this catalog are numerous. Measurements of statistical properties of the cluster population, such as the cluster-cluster correlation function $\xi_{cc}$ and the cluster mass function are all avenues of further research. Comparisons to the relatively large X-ray selected samples from the RASS are forthcoming, with the potential to find many poorer systems by performing joint optical/X-ray detections. The existence of independent samples of thousands of clusters will allow us to evaluate the biases present in the different detection methods.

We also plan to examine the multiplicity function \citep{pud03}, from small groups to rich clusters. This requires a separate catalog of poor groups, which is being constructed using a different algorithm by \citep{iov03}. The catalog presented here and the compact group catalog are expected to have significant overlap, which will also provide important cross-checks in the difficult domain of small galaxy associations.

This sample is also fertile ground for the selection of specific, well-defined subsamples for follow-up observational studies. Using these cluster locations and the now public DPOSS data, we can search for clusters with substructure, excess blue galaxy populations, unusual optical/X-ray flux ratios, or any of a variety of interesting properties. Other methods for cluster detection have also been applied to our galaxy catalogs \citep{pud01}, which can be used for further testing and comparisons.

We remind the reader that all of the data for this survey (including future installments) can be found at \texttt{http://dposs.caltech.edu/dataproducts/}.

\acknowledgments

We thank the Norris Foundation and other private donors for their generous support of the DPOSS project. RRG was supported in part by an NSF Fellowship, NASA GSRP NGT5-50215, a Kingsley Fellowship, as well as a KDI grant to A. Szalay. This work would have been impossible without the POSS-II photographic team and the STScI digitization team. We also thank the Palomar TAC and Directors for generous time allocations for the DPOSS calibration effort. The DPOSS survey would have been impossible without the hard work of the POSS-II observing staff, and numerous undergraduates who assisted with the calibration. RRG would like to thank Sandra Castro for assistance with the spectroscopy reductions, and RdC thanks Hugo Capelato, Angela Iovino and Gary Mamon for discussion on several issues in this project. This work was made possible in part through the NPACI sponsored Digital Sky project and a generous equipment grant from SUN Microsystems. Access to the POSS-II image data stored on the HPSS, located at the California Institute of Technology, was provided by the Center for Advanced Computing Research.

\clearpage

\begin{figure}
\epsscale{0.9}
\plotone{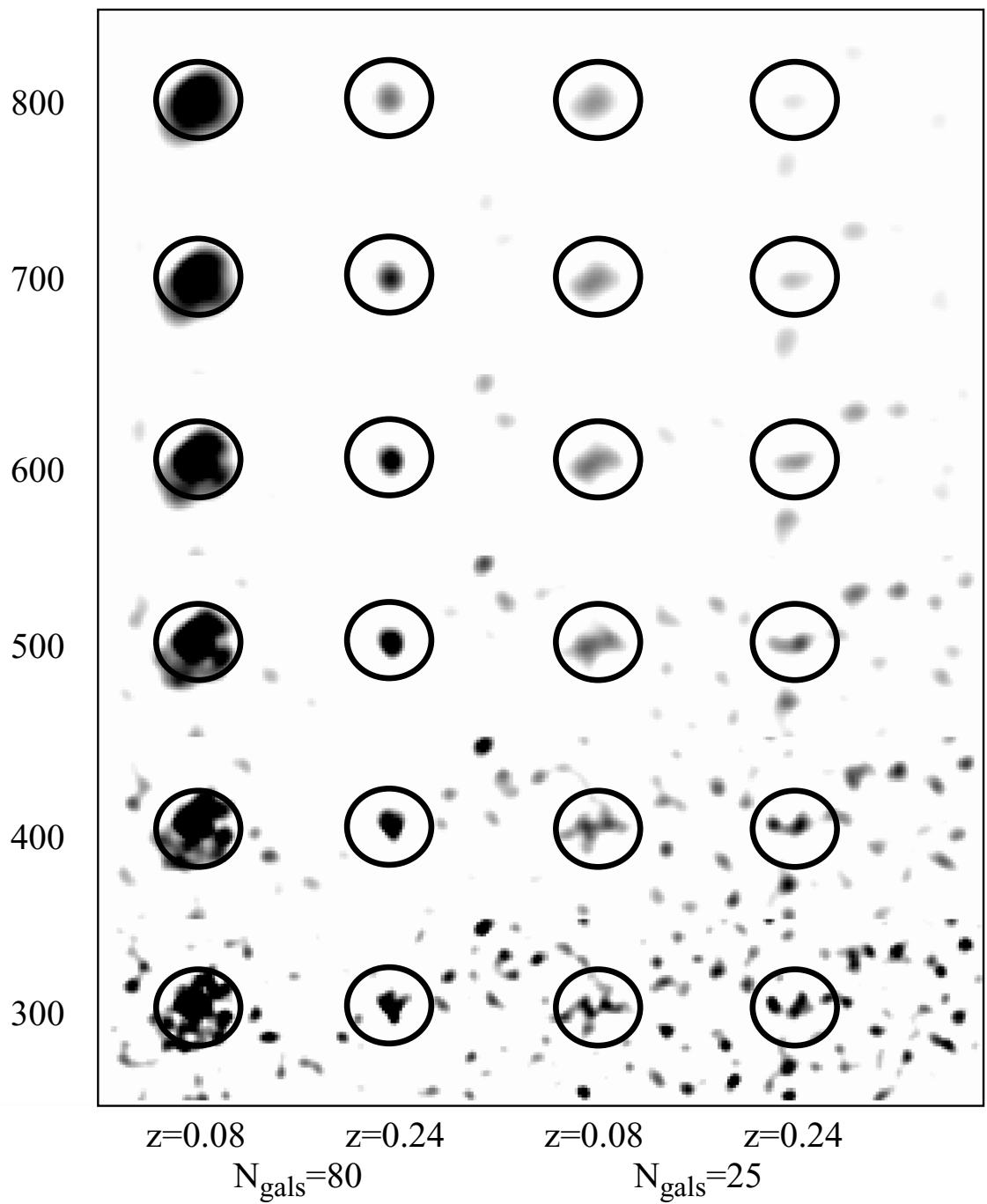}
\caption{The effect of varying the initial smoothing window on cluster appearance. Each panel contains a simulated background with four simulated clusters, as described in the text. The smoothing kernel ranges in size from $300''$ for the upper panel, to $800''$ for the lower panel, in $100''$ increments.\label{fig1}}
\end{figure}

\begin{figure}
\plotone{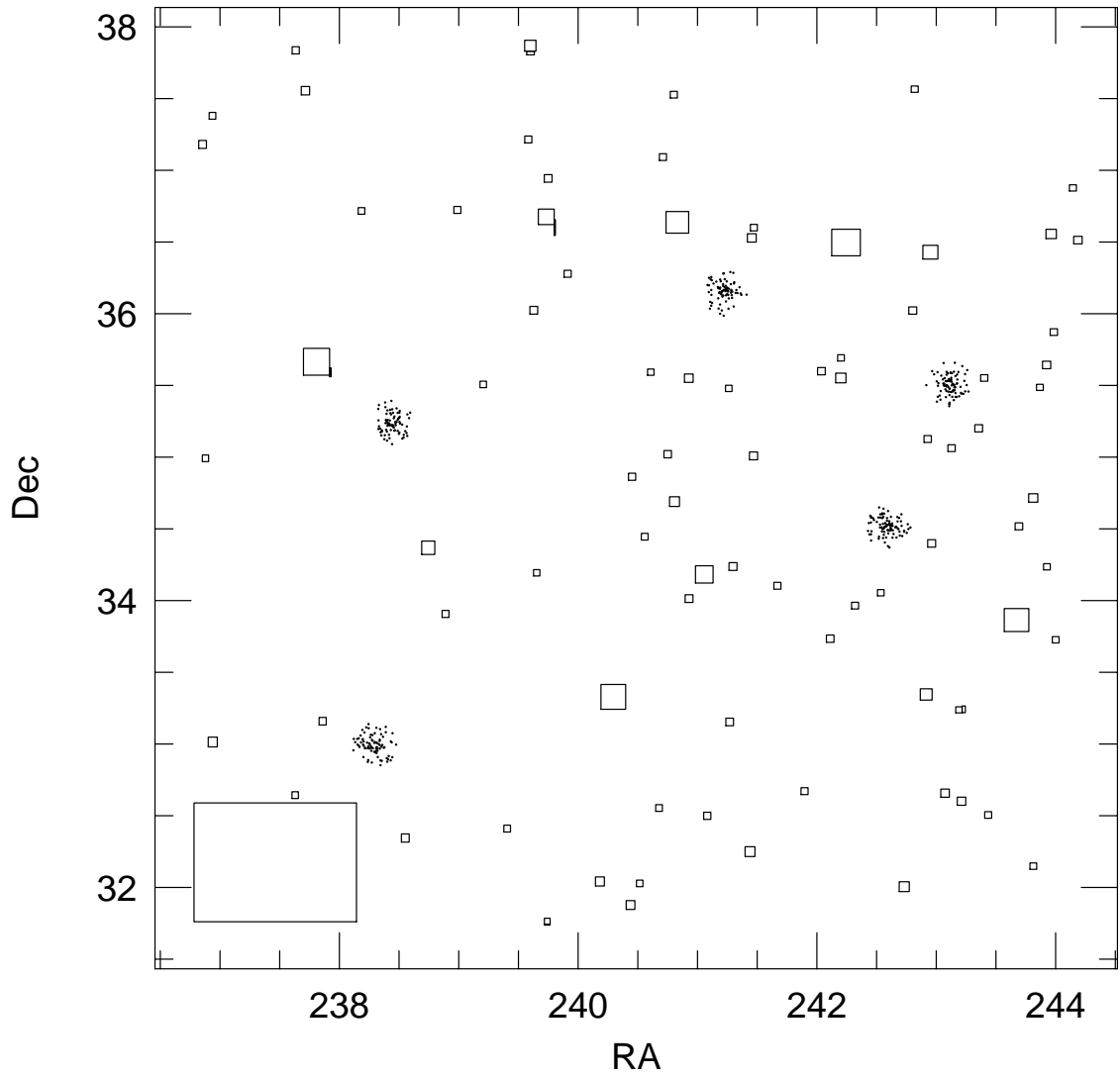}
\caption{Plate 389, with the bad areas marked, and five simulated clusters at $z=0.16$ and $N_{gals}=80$ inserted.\label{fig2}}
\end{figure}

\begin{figure}
\epsscale{0.7}
\plotone{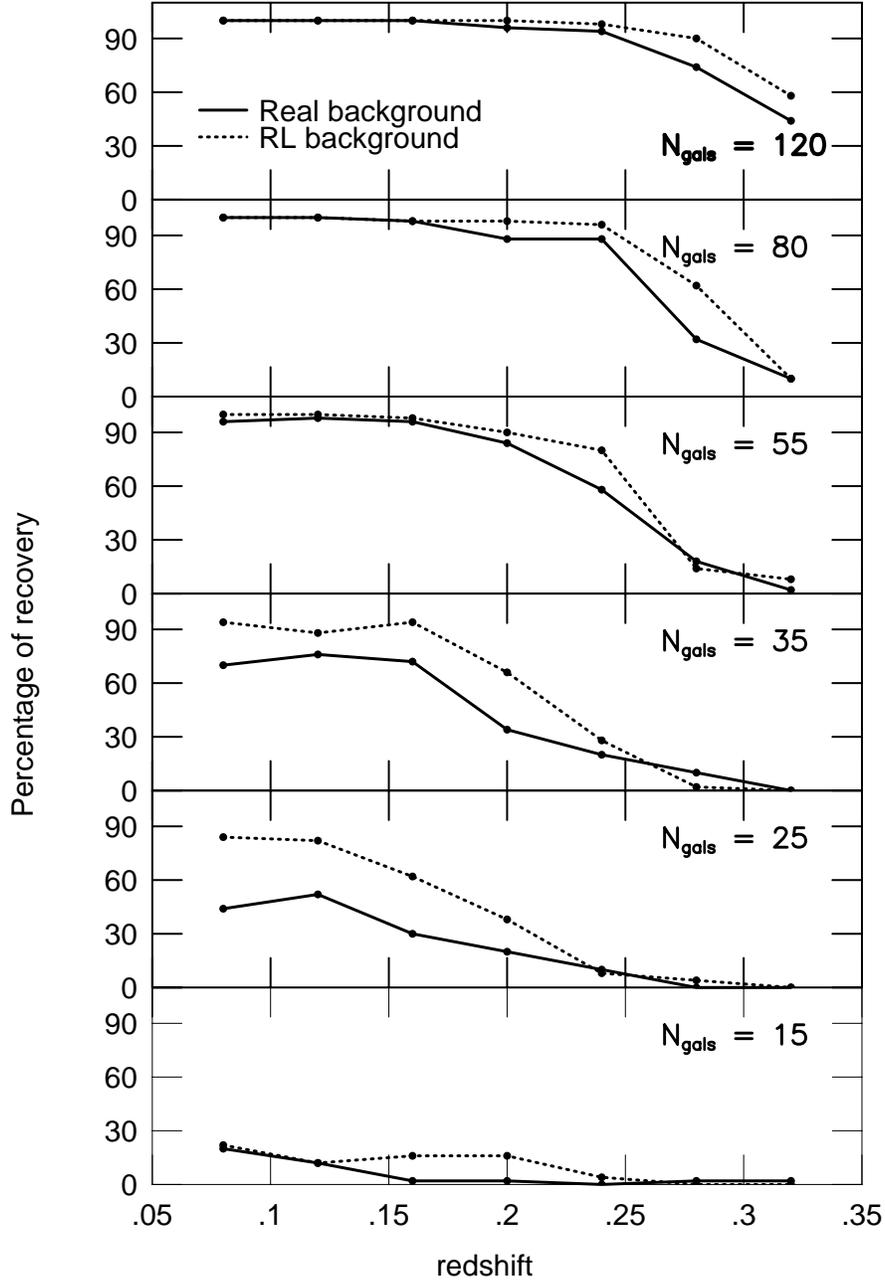}
\caption{The completeness rate evaluated with two different backgrounds:
a RL distribution (dashed line) and a real galaxy catalog from DPOSS 
(full line). The richness, increasing from bottom to top,
is indicated on each panel.\label{fig3}}
\end{figure}

\begin{figure}
\plotone{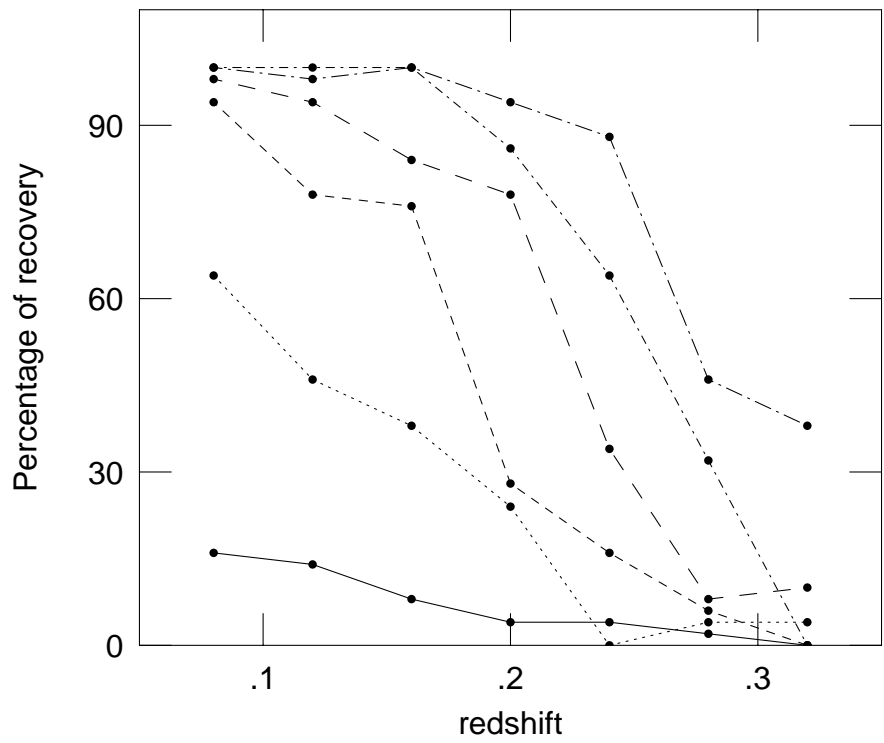}
\caption{An example completeness function, for Plate 389. From top to bottom, the functions correspond to clusters of richnesses $N_{gals}=[120,80,55,35,25,15]$.\label{fig4}}
\end{figure}

\begin{figure}
\epsscale{0.85}
\plotone{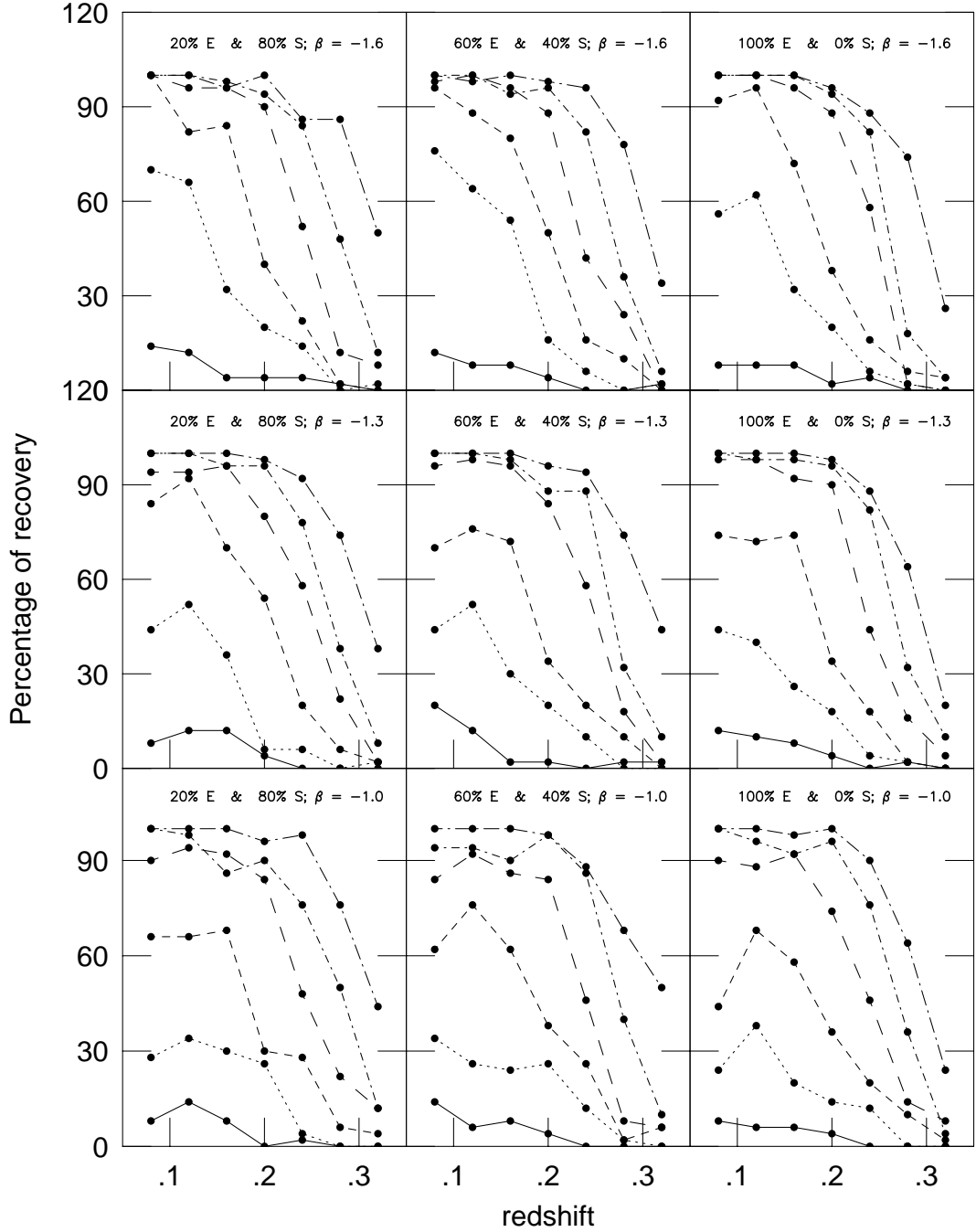}
\caption{The selection function evaluated with different cluster compositions
and spatial profile slopes $\beta$. From bottom to top $\beta$ assumes
the values -1.0, -1.3 and -1.6. The cluster composition, from right
to left panels is 20\% E + 80\% Sbc, 60\% E + 40\% Sbc, and 100\% E. Richness classes are the same as Figure 4.\label{fig5}}
\end{figure}

\begin{figure}
\epsscale{0.85}
\plotone{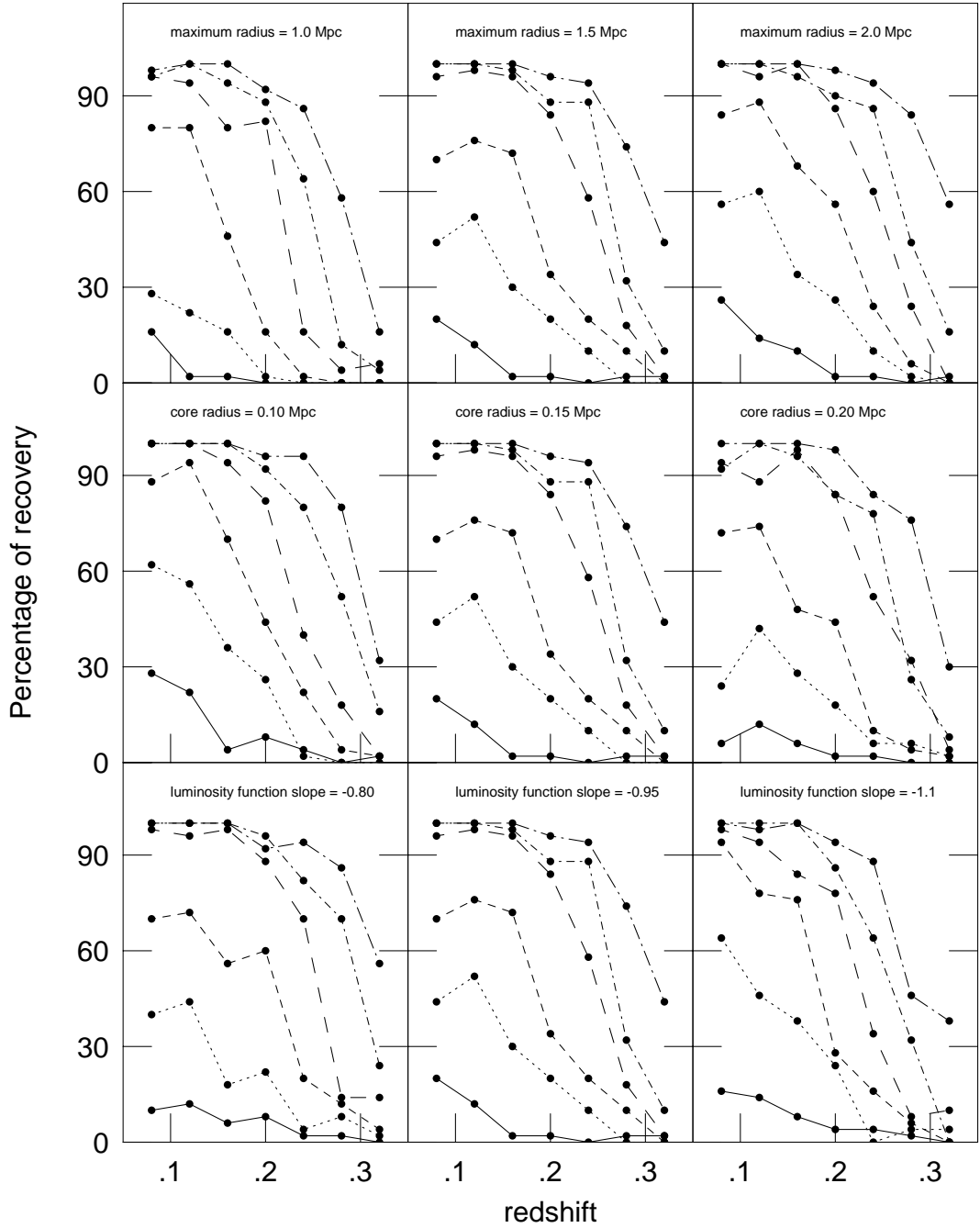}
\caption{Dependence of the selection function with the
luminosity function slope (bottom panels), core radius
(middle panels) and cutoff radius (top panels).  The values tested, from left to right, are  $\alpha = -0.8, -0.95, -1.1$ for the LF slope, 
$r_{core} = 0.1, 0.15, 0.2 h^{-1}$ Mpc for the core radius, and
$r_{max} = 1.0, 1.5, 2.0$ $h^{-1}$ Mpc for the maximum radius. Richness classes are the same as shown in Figure 4.\label{fig6}}
\end{figure}

\begin{figure}
\plotone{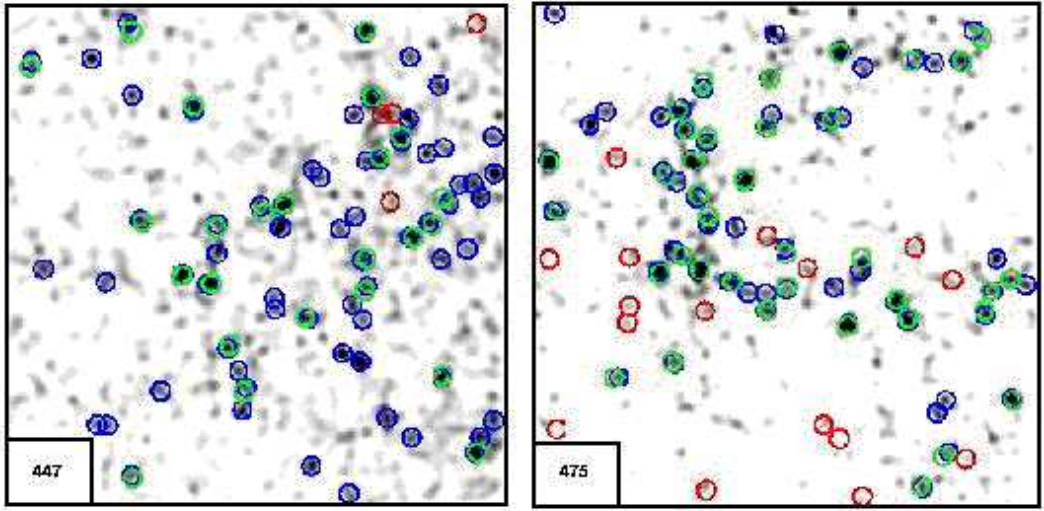}
\caption{AK galaxy density maps for Fields 447 (left) and 475 (right), with blue circles marking the locations of current candidate clusters, green circles showing slitmasks associated with current candidates, and red circles showing locations where spectra were taken but there is no current candidate. \label{fig7}}
\end{figure}

\begin{figure}
\plotone{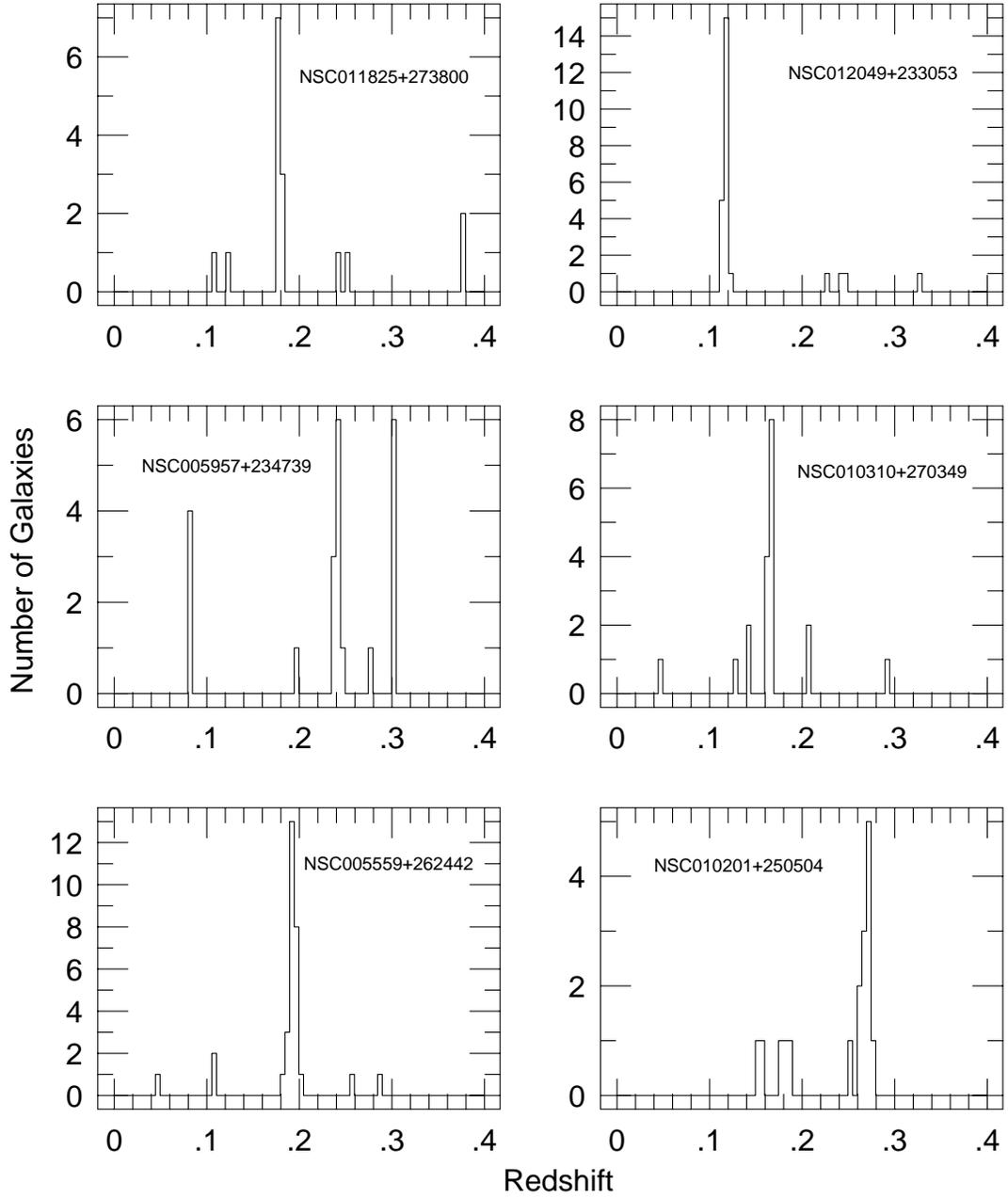}
\caption{Spectroscopic redshift histograms for selected candidates in Field 475.\label{fig8}}
\end{figure}

\begin{figure}
\epsscale{0.8}
\plotone{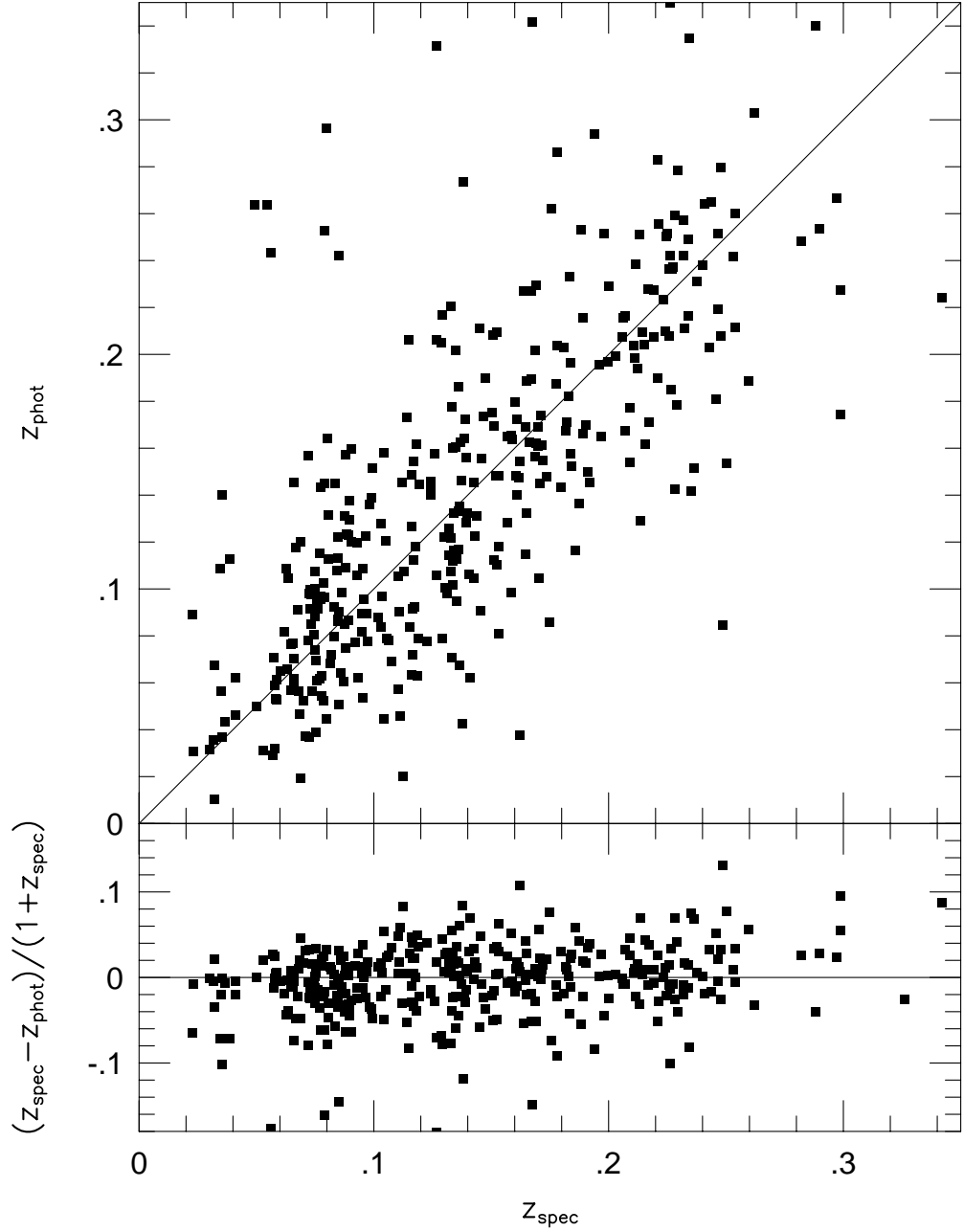}
\caption{The photometrically estimated redshift {\em vs.} the spectroscopically measured redshift for 369 Abell clusters. Residuals as a function of redshift are shown in the bottom panel.\label{fig9}}
\end{figure}

\begin{figure}
\plotone{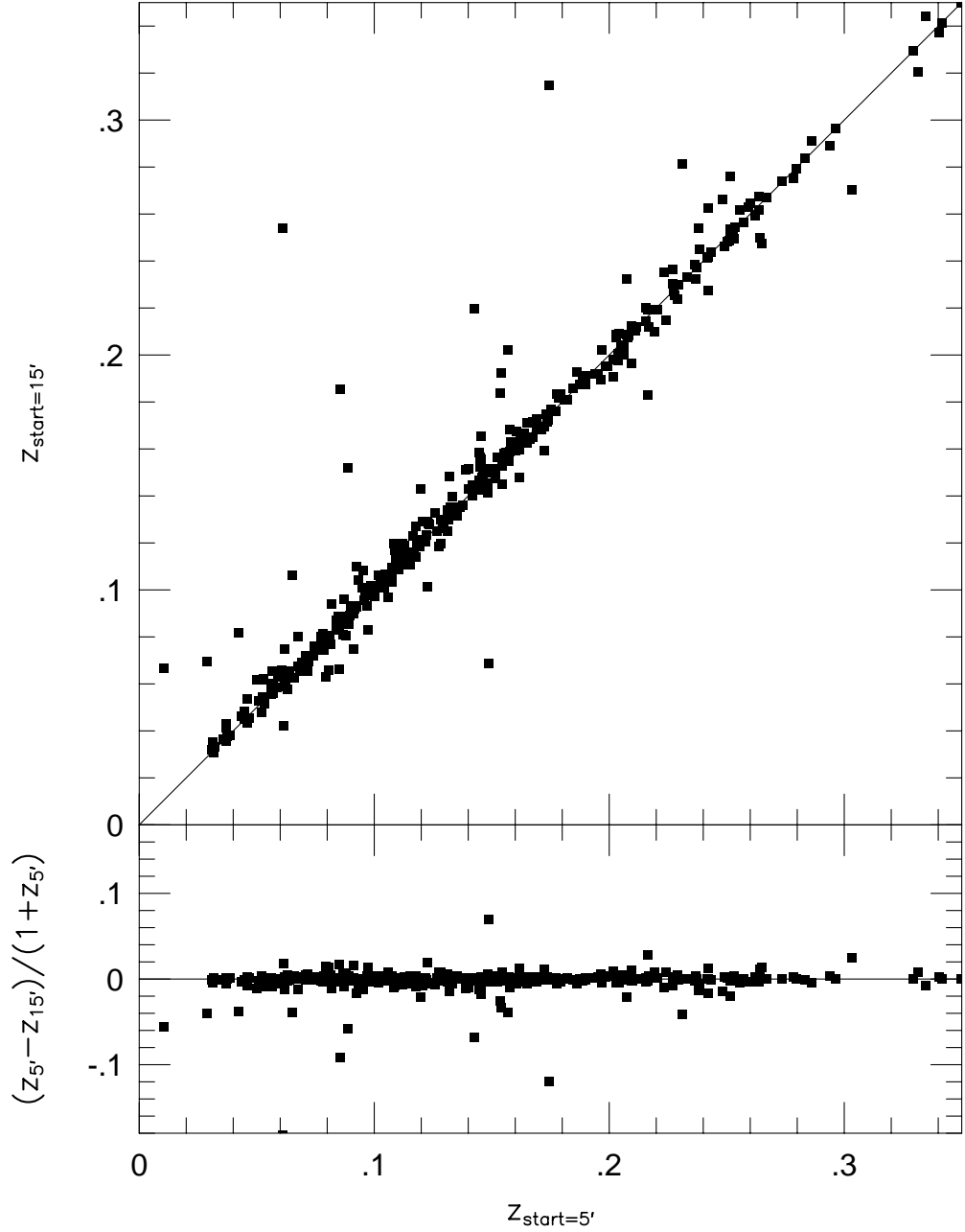}
\caption{The photometrically estimated redshift using starting radii of 5'and 15'. Differences between the estimates are shown in the bottom panel.\label{fig10}}
\end{figure}

\begin{figure}
\plotone{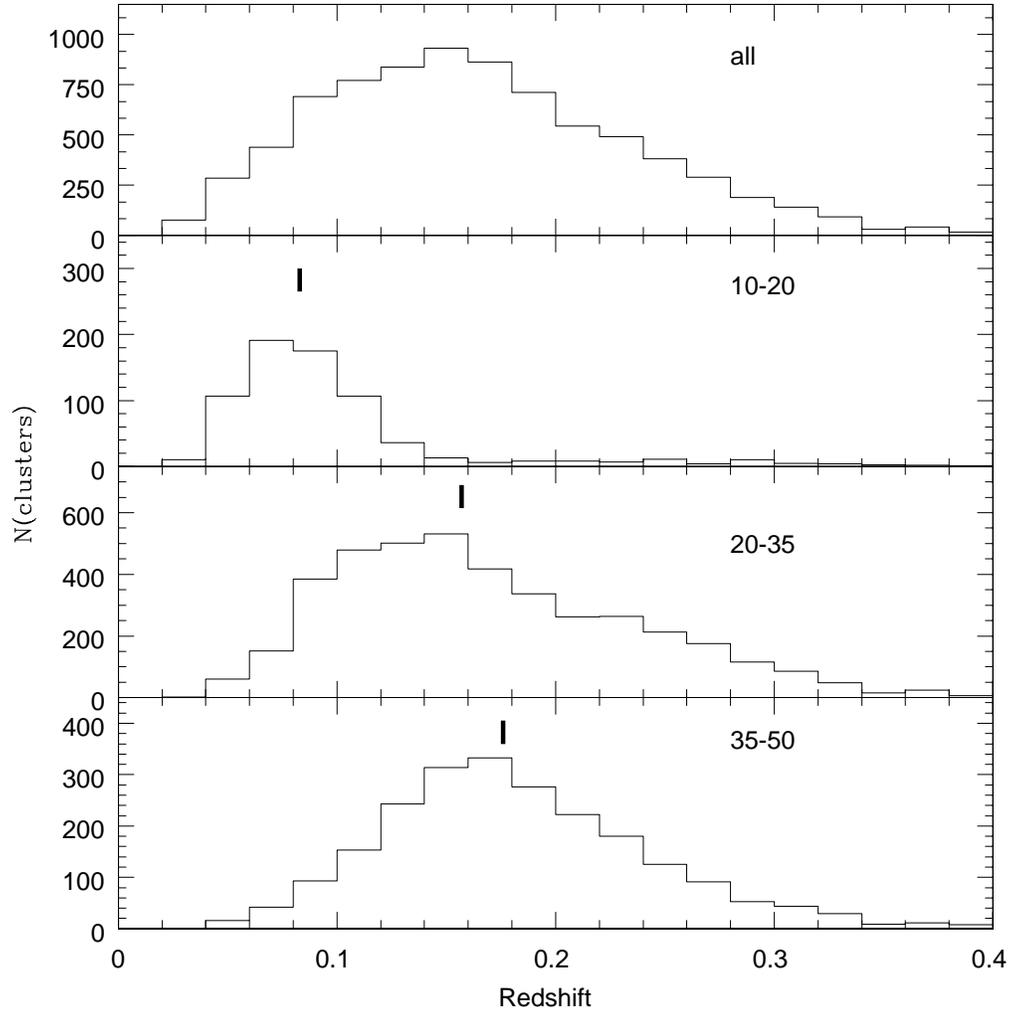}
\caption{The redshift distribution of our candidate clusters. The distribution for all clusters is shown at top, with lower panels dividing the sample by $N_{gals}$.\label{fig11}}
\end{figure}

\begin{figure}
\plotone{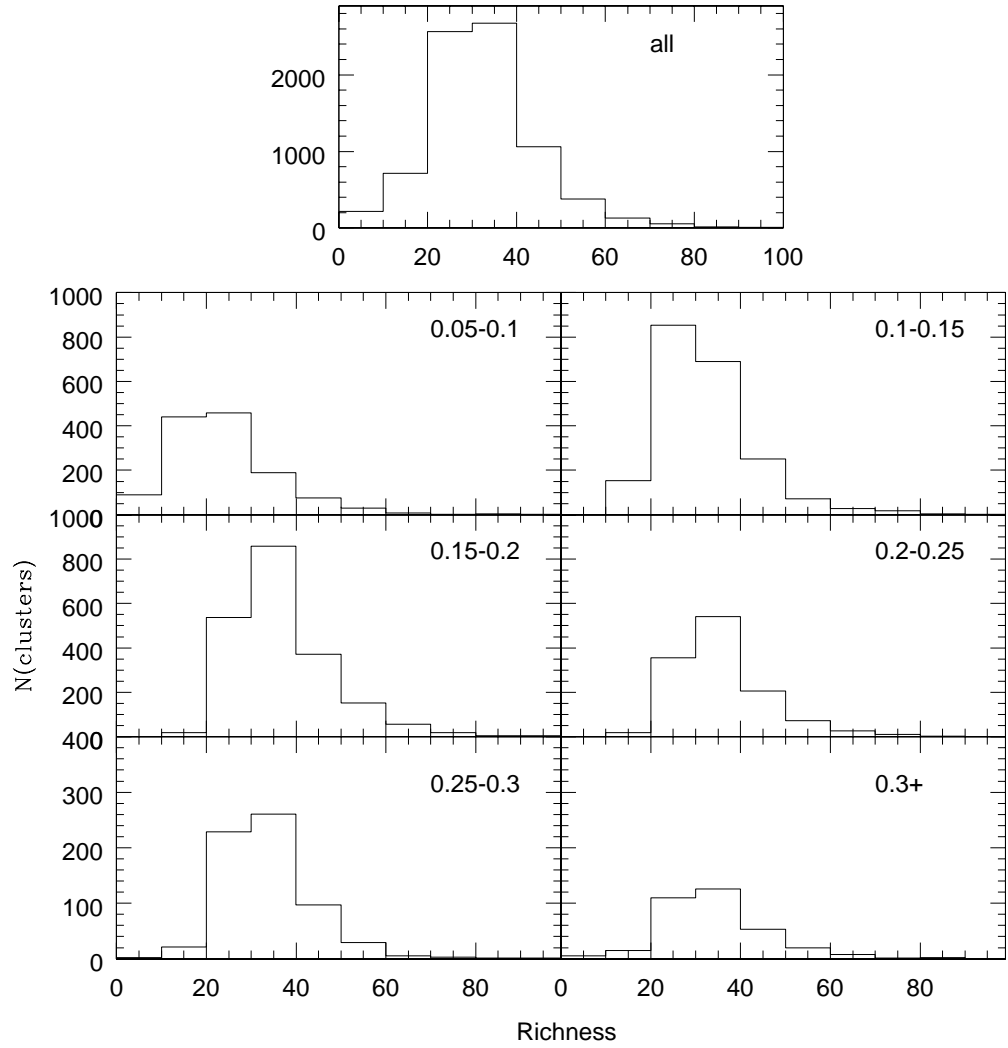}
\caption{Richness distributions for the entire sample (top panel) and in different redshift bins (lower panels).\label{fig12}}
\end{figure}

\begin{figure}
\epsscale{1.0}
\plotone{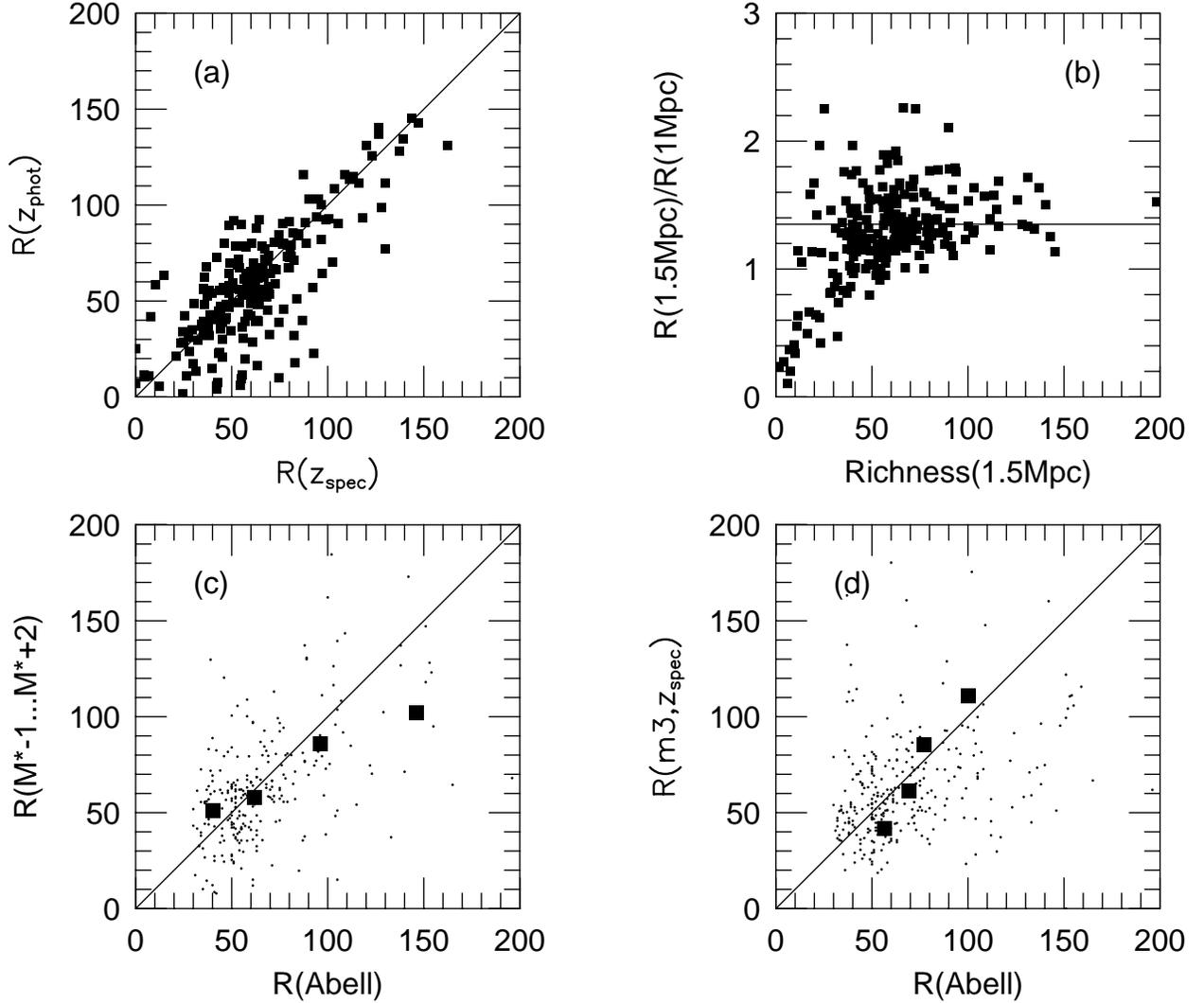}
\caption{Richness estimator tests using Abell clusters with spectroscopic redshifts. Panel (a) shows our richness measure using the spectroscopic redshift from SR99 compared to the richness using our photo-$z$. Panel (b) is the ratio of richnesses using 1.5Mpc and 1Mpc radii. Panel (c) shows our richness (using $z_{spec}$) against Abell's. Panel (d) compares Abell's richness to our measurement using the \citet{kim01} technique to count galaxies with $m_3<m<m_3+2$.In the lower panels, each small dot is a single cluster, with the large squares representing binned data.
\label{fig13}}
\end{figure}

\begin{figure}
\plotone{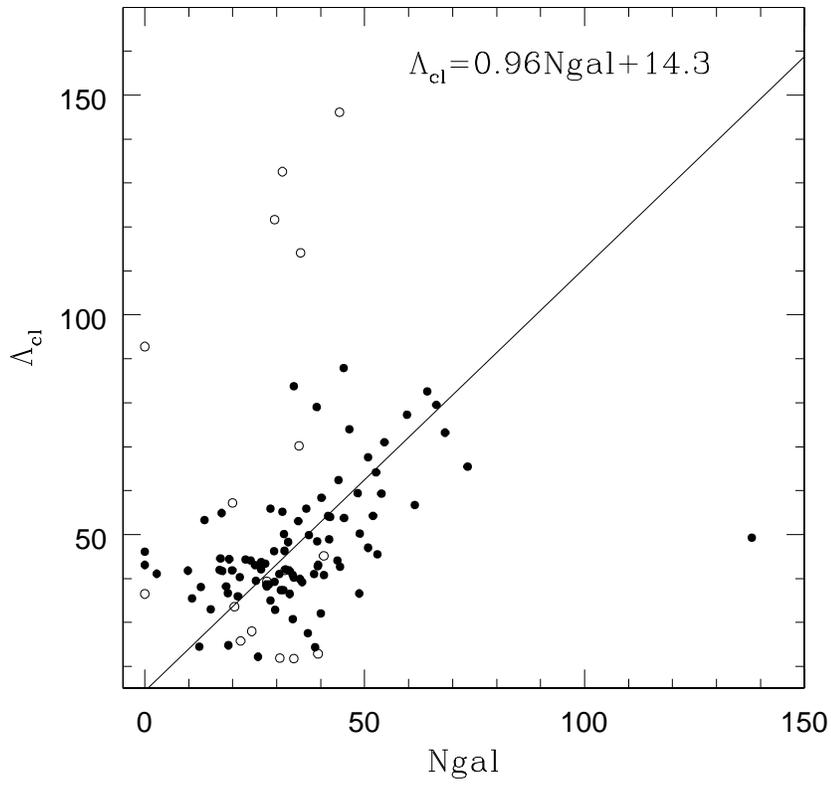}
\caption{Comparison of our richness measure, $N_{gals}$, with independent measurements of $\Lambda_{cl}$ from \citet{kim01}.\label{fig14}}
\end{figure}

\begin{figure}
\plotone{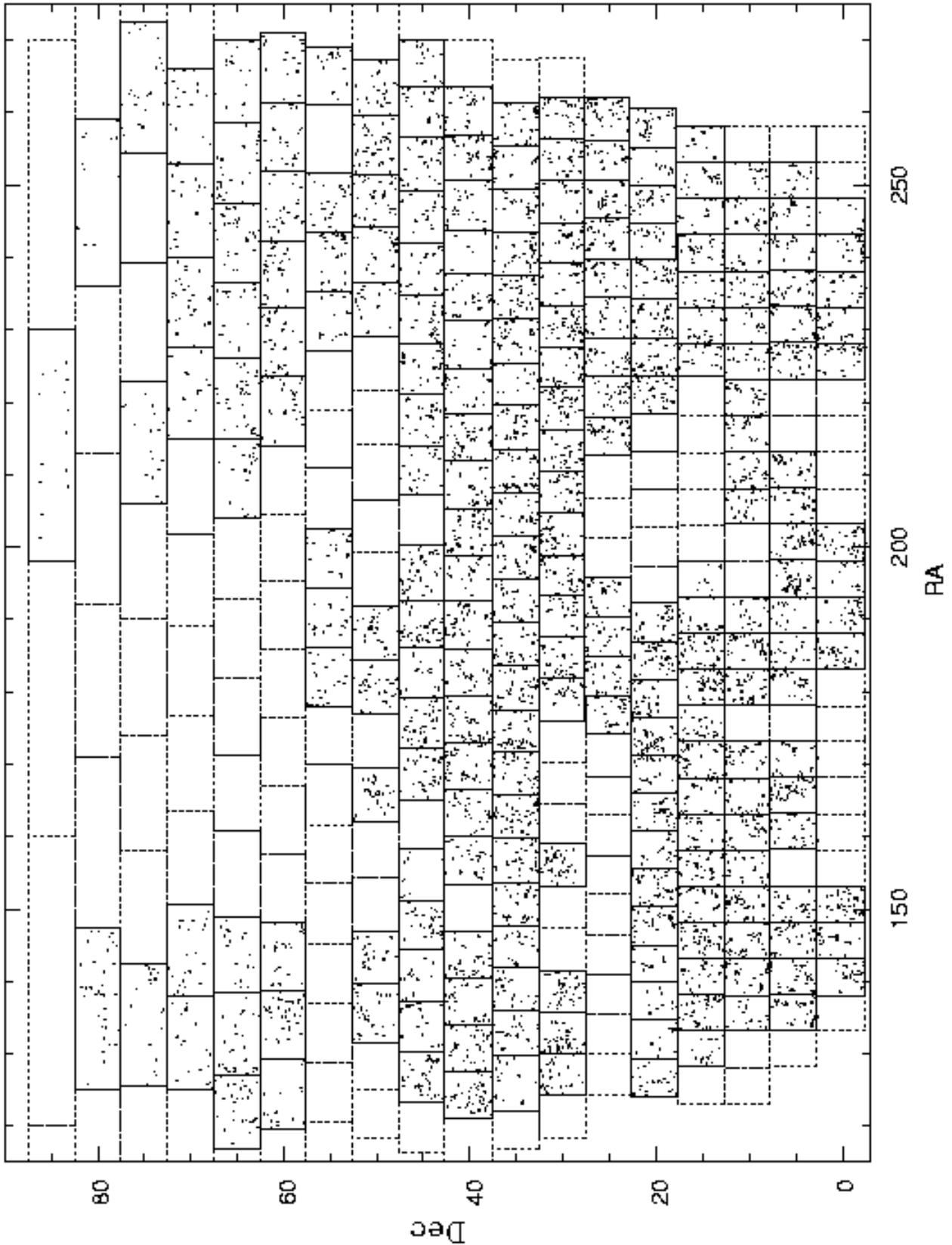}
\caption{The sky distribution of our candidate clusters. Black lines show the boundaries of plates used in this paper; dotted boundaries are unused plates with poorer calibration.\label{fig15}}
\end{figure}

\begin{figure}
\plotone{Gal.clusters.fig16.epsi}
\caption{DPOSS $F$-plate images of four rich clusters. Two were not previously known, and two (NSC172013+264028 = RXC J1720.1+2637 and  NSC122906+473720 = RXC J1229.0+4737) were detected only as X-ray clusters. \label{fig16}}
\end{figure}

\clearpage

\begin{deluxetable}{cccc}
\tablecolumns{4} 
\tablewidth{0pc} 
\tablecaption{Percentage of Clusters Detected vs. Number of Maps \label{tbl-1}}
\tablehead{
\colhead{Percent} & \colhead{N(maps)} & \colhead{Percent} & \colhead{N(maps)}}
\startdata
 99.9 & 1 & 92.8 & 6 \\
 99.7 & 2 & 89.7 & 7 \\
 98.9 & 3 & 68.6 & 8 \\
 98.5 & 4 & 66.3 & 9 \\
 96.3 & 5 & 62.1 & 10 \\
\enddata
\end{deluxetable}

\begin{deluxetable}{ccccccccc}
\tablecolumns{8} 
\tablewidth{0pc} 
\tablecaption{Example Completeness Function: Plate 389}
\tablehead{
\colhead{$\mathbf{N/z}$} & \colhead{\bf{0.08}} & \colhead{\bf{0.12}} & \colhead{\bf{0.16}} & \colhead{\bf{0.20}} & \colhead{\bf{0.24}}  & \colhead{\bf{0.28}} & \colhead{\bf{0.32}}}
\startdata 
{\bf 15} & 0.16 & 0.14 & 0.08 & 0.04 & 0.04 & 0.02 & 0.00 \\
{\bf 25} & 0.64 & 0.46 & 0.38 & 0.24 & 0.00 & 0.04 & 0.04 \\
{\bf 35} & 0.94 & 0.78 & 0.76 & 0.28 & 0.16 & 0.06 & 0.00 \\
{\bf 55} & 0.98 & 0.94 & 0.84 & 0.78 & 0.34 & 0.08 & 0.10 \\
{\bf 80} & 1.00 & 1.00 & 1.00 & 0.86 & 0.64 & 0.32 & 0.00 \\
{\bf 120} & 1.00 & 0.98 & 1.00 & 0.94 & 0.88 & 0.46 & 0.38 \\
\enddata 
\end{deluxetable} 

\begin{deluxetable}{cccr} 
\tablecolumns{4} 
\tablewidth{0pc} 
\tablecaption{Spectroscopic Survey Results}
\tablehead{
\colhead{Field} & \colhead{N(cands)} & \colhead{N(spec) [frac]} & \colhead{N(extra)} }
\startdata 
447 & 64 & 24 [0.38] & 4 \\
475 & 55 & 37 [0.67] & 16 \\
\enddata 
\end{deluxetable} 

\begin{deluxetable}{ccrc} 
\tablecolumns{4} 
\tablewidth{0pc} 
\tabletypesize{\small}
\tablecaption{Field 447 Spectroscopy} 
\tablehead{ 
\colhead{Candidate/Mask}  & \colhead{$z_{spec}$} & \colhead{$N(z)$} & \colhead{Comment} }
\startdata 
NSC142311+320840 & 0.200 & 10 &  \\
NSC142841+323859 & 0.127 & 5 &  \\
NSC142920+270609 & 0.268 & 6 &  \\
NSC142937+301403 & 0.103 & 4 &  \\
NSC143203+293404 & 0.221 & 14 &  \\
NSC143237+313532 & 0.131 & 13 &  \\
NSC143330+292738 & 0.219 & 12 &  \\
NSC143400+301222 & 0.222 & 3 &  \\
NSC143437+284024 & 0.205 & 11 &  \\
NSC143539+281143 & 0.203 & 3 &  \\
NSC143737+300923 & 0.338 & 9 &  \\
NSC143744+302547 & 0.160 & 13 &  \\
NSC143910+290229 & 0.253 & 11 &  \\
NSC144210+294444 & 0.216 & 4 &  \\
NSC144229+292545 & 0.224 & 8 &  \\
NSC144231+323227 & 0.243 & 11 &  \\
NSC144250+314342 & 0.243 & 9 &  \\
NSC144315+305758 & 0.227 & 4 &  \\
NSC144432+311149 & 0.233 & 14 &  \\
NSC144457+300112 & 0.177 & 3 &  \\
NSC144603+301148 & 0.109 & 5 &  \\
NSC144635+281740 & 0.229 & 4 &  \\
NSC144713+302554 & 0.170 & 4 &  \\
NSC144820+272134 & 0.233 & 20 &  \\
144328+313136 & 0.240 & 11 & High $z$; $N_{gals}=36.3$ \\
144352+302724 & 0.320 & 4 & High $z$; $N_{gals}=22.5$ \\
144404+313214 & 0.233 & 6 & High $z$; $N_{gals}=25.3$ \\
144902+323713 & 0.200 & 3 & $N_{gals}=13.2$ \\
\enddata 
\end{deluxetable}

\begin{deluxetable}{ccrc} 
\tablecolumns{4} 
\tablewidth{0pc} 
\tabletypesize{\small}
\tableheadfrac{0}
\tablecaption{Field 475 Spectroscopy} 
\tablehead{ 
\colhead{Candidate/Mask}  & \colhead{$z_{spec}$} & \colhead{$N(z)$} & \colhead{Comment} }
\startdata 
NSC005559+262442 & .194 & 24 &  \\
NSC005618+254729 & .150/.245 & 3/3 &  \\
NSC005957+234739 & .240/.302 & 10/6 &  \\
NSC010201+250504 & .273 & 11 &  \\
NSC010211+261816 & .239 & 7 &  \\
NSC010251+252028 & .189 & 15 &  \\
NSC010255+235859 & .266 & 6 &  \\
NSC010310+270349 & .166 & 12 &  \\
NSC010319+264850 & .127:: & 3 &  \\
NSC010348+262628 & .242 & 12 &  \\
NSC010403+255906 & .245 & 7 &  \\
NSC010408+250654 & .160 & 17 &  \\
NSC010420+271828 & .239 & 8 &  \\
NSC010434+263908 & .168 & 7 &  \\
NSC010439+254013 & .158 & 15 &  \\
NSC010546+245803 & .241 & 17 &  \\
NSC010641+261142 & .164 & 15 &  \\
NSC010748+243721 & .238 & 7 &  \\
NSC010749+265059 & .192 & 8 &  \\
NSC010758+272626 & .117/.238 & 4/3 &  \\
NSC010847+252214 & .200 & 12 &  \\
NSC010853+245311 & .197 & 10 &  \\
NSC011059+265458 & .113 & 7 &  \\
NSC011152+274612 & .115 & 8 &  \\
NSC011210+242646 & .196 & 17 &  \\
NSC011251+250601 & .183 & 5 &  \\
NSC011444+244250 & .178 & 8 &  \\
NSC011521+242925 & .189 & 6 &  \\
NSC011601+222736 & .248 & 7 &  \\
NSC011601+273938 & .120 & 8 &  \\
NSC011726+225257 & .123 & 6 &  \\
NSC011825+273800 & .177 & 10 &  \\
NSC011932+243213 & .142 & 7 &  \\
NSC011954+244954 & .207 & 5 &  \\
NSC012049+233053 & .117 & 21 &  \\
NSC012057+245751 & .190 & 9 &  \\
010030+243949 & .082 & 7 & $N_{gals}=17.1$ \\
011252+222124 & .140 & 6 & $N_{gals}=16.0$ \\
010432+243635 & .266 & 9 & $N_{gals}=20.4$ \\
005940+262717 & .193 & 14 & $N_{gals}=22.8$ \\
005645+230911 & .180 & 6 & $N_{gals}=18.9$ \\
010441+222618 & .250 & 5 & High $z$; $N_{gals}=22.9$ \\
010752+253143 & .200 & 6 & $N_{gals}=29.4$ \\
011142+230337 & .195 & 7 & $N_{gals}=9.8$ \\
011818+224753 & .267 & 5 & High $z$; $N_{gals}=27.63$ \\
010025+242744 & .125 & 10 & $N_{gals}=20.3$ \\
011000+250756 & .200 & 6 & $N_{gals}=26.7$ \\
011050+231344 & .115 & 5 & $N_{gals}=23.4$ \\
010027+251510 & .227 & 5 & High $z$; $N_{gals}=28.6$  \\
011548+252229 & .185 & 9 & $N_{gals}=22.8$ \\
011747+245810 & .145 & 11 & $N_{gals}=23.2$ \\
005608+251240 & .175 & 8 & $N_{gals}=16.80$ \\
\enddata 
\end{deluxetable} 
 
\begin{deluxetable}{crrlcrc} 
\tabletypesize{\footnotesize}
\tablecolumns{7} 
\tablewidth{0pc} 
\tablecaption{The Northern Sky Cluster Catalog: Excerpt}
\tablehead{
\colhead{Name} & \colhead{RA (J2000.0)} & \colhead{Dec (J2000.0)} & \colhead{$z_{phot}$} & \colhead{$N_{gals}$} & \colhead{$N_{maps}$} & \colhead{Plate}
}
\startdata 
NSC080140+623236  & 120.41708 &  62.54339 &  0.0732 &  28.1 & 10 & 124 \\ 
NSC080142+575020  & 120.42686 &  57.83877 &  0.1350 &  20.6 &  7 & 124 \\ 
NSC080205+634700  & 120.52274 &  63.78326 &  0.0725 &  45.9 & 10 & 089 \\ 
NSC080217+633240  & 120.57254 &  63.54455 &  0.0626 &  22.9 & 10 & 089 \\ 
NSC080218+604112  & 120.57571 &  60.68660 &  0.1080 &  23.6 & 10 & 124 \\ 
NSC080244+642523  & 120.68704 &  64.42318 &  0.1107 &  28.6 &  7 & 089 \\ 
NSC080245+612414  & 120.69027 &  61.40387 & 0.1149: &  29.8 & 10 & 124 \\ 
NSC080300+631419  & 120.75366 &  63.23854 &  0.1434 &  38.3 &  8 & 089 \\ 
NSC080320+652141  & 120.83540 &  65.36151 &  0.1470 &  29.5 &  7 & 089 \\ 
NSC080327+640531  & 120.86494 &  64.09192 &  0.1341 &  28.6 &  7 & 089 \\ 
NSC080337+585755  & 120.90653 &  58.96515 &  0.2260 &  23.4 & 10 & 124 \\ 
NSC080347+630320  & 120.94792 &  63.05546 &  0.0874 &  30.0 &  9 & 089 \\ 
NSC080401+625402  & 121.00716 &  62.90060 &  0.0850 &  33.9 &  8 & 089 \\ 
NSC080435+395527  & 121.14706 &  39.92425 &  0.1014 &  27.7 & 10 & 312 \\ 
NSC080542+670818  & 121.42574 &  67.13827 &  0.0942 &  32.2 & 10 & 089 \\ 
NSC080608+374711  & 121.53357 &  37.78649 & 0.2323: &  30.2 & 10 & 312 \\ 
NSC080615+381832  & 121.56293 &  38.30891 & 0.2619: &  60.7 & 10 & 312 \\ 
NSC080621+614129  & 121.58971 &  61.69127 &  0.1691 &  33.1 & 10 & 124 \\ 
NSC080650+620358  & 121.71079 &  62.06607 &  0.2518 &  37.6 & 10 & 124 \\ 
NSC080717+393521  & 121.82310 &  39.58903 &  0.0936 &  18.6 & 10 & 312 \\ 
NSC080745+605029  & 121.94125 &  60.84137 &  0.2797 &  29.7 & 10 & 124 \\ 
\enddata 
\end{deluxetable}

\end{document}